\documentclass[letterpaper]{article}
\usepackage[preprint]{aaai2027}
\usepackage[hyphens]{url}
\usepackage{graphicx}
\usepackage{natbib}
\usepackage{caption}
\usepackage{amsmath,amssymb,bm}
\usepackage{booktabs}
\usepackage{makecell}
\usepackage{multirow}

\graphicspath{{Figures/}}

\title{PACT: Phenotype-Aware Contrastive Team Representation\\for Multi-Phenotype Grouped Ad Hoc Teamwork}
\author{
    Beiwen Zhang\textsuperscript{\rm 1}\equalcontrib,
    Yongheng Liang\textsuperscript{\rm 1}\equalcontrib,
    Guowei Zou\textsuperscript{\rm 1},
    Haitao Wang\textsuperscript{\rm 1},
    Cong Liu\textsuperscript{\rm 1},
    Hejun Wu\textsuperscript{\rm 1}\corresponding
}
\affiliations{
    \textsuperscript{\rm 1}School of Computer Science and Engineering, Sun Yat-sen University\\
    Guangzhou, China\\
    zhangbw39@mails.sysu.edu.cn, liangyh38@mail2.sysu.edu.cn, zougw@mail2.sysu.edu.cn,\\
    wanght76@mail.sysu.edu.cn, liucong3@mail.sysu.edu.cn, wuhejun@mail.sysu.edu.cn
}

\begin{document}

\maketitle

\begin{abstract}
Learning to collaborate with various unfamiliar teammates poses a great challenge in the domain of multi-agent systems. Existing ad hoc teamwork methods typically drive controlled agents to collaborate with a group of teammates exhibiting a single coordination phenotype shaped by the same reward function. However, in real-world applications, controlled agents should collaborate with unfamiliar teammates of diverse coordination phenotypes among groups that have never worked together. We formalize this as the \textit{Multi-Phenotype Grouped Ad Hoc Teamwork (MPG-AHT)} problem, and propose \textit{Phenotype-Aware Contrastive Team Representation (PACT)} to solve this problem. PACT is empowered with phenotype-aware contrastive learning and relational reasoning to accurately distinguish coordination phenotypes and capture inter-agent interactions. Extensive experiments on multi-phenotype collaboration tasks show that PACT outperforms state-of-the-art baselines on average, achieving a mean 21.0\% gain in out-of-distribution evaluation and a mean 36.5\% gain in sample efficiency. The appendix is included in this version.
\end{abstract}

\section{Introduction}
Multi-agent reinforcement learning (MARL) has shown strong performance in collaborative tasks across domains such as autonomous driving \cite{driving}, robotic systems \cite{MARL_robot}, and distributed resource management \cite{distributed}. 
Most MARL studies assume that agents are trained within a fixed team and jointly controlled, allowing coordination to emerge through repeated interaction with the same teammates \cite{MARL_example1}. 
To enable collaboration in more open multi-agent environments, the ad hoc teamwork (AHT) paradigm \cite{FirstAht} studies how an agent under control collaborates with unfamiliar teammates beyond control whose policies are inaccessible to the learner.

\begin{figure}[t]
	\centering
	\includegraphics[width=1\columnwidth]{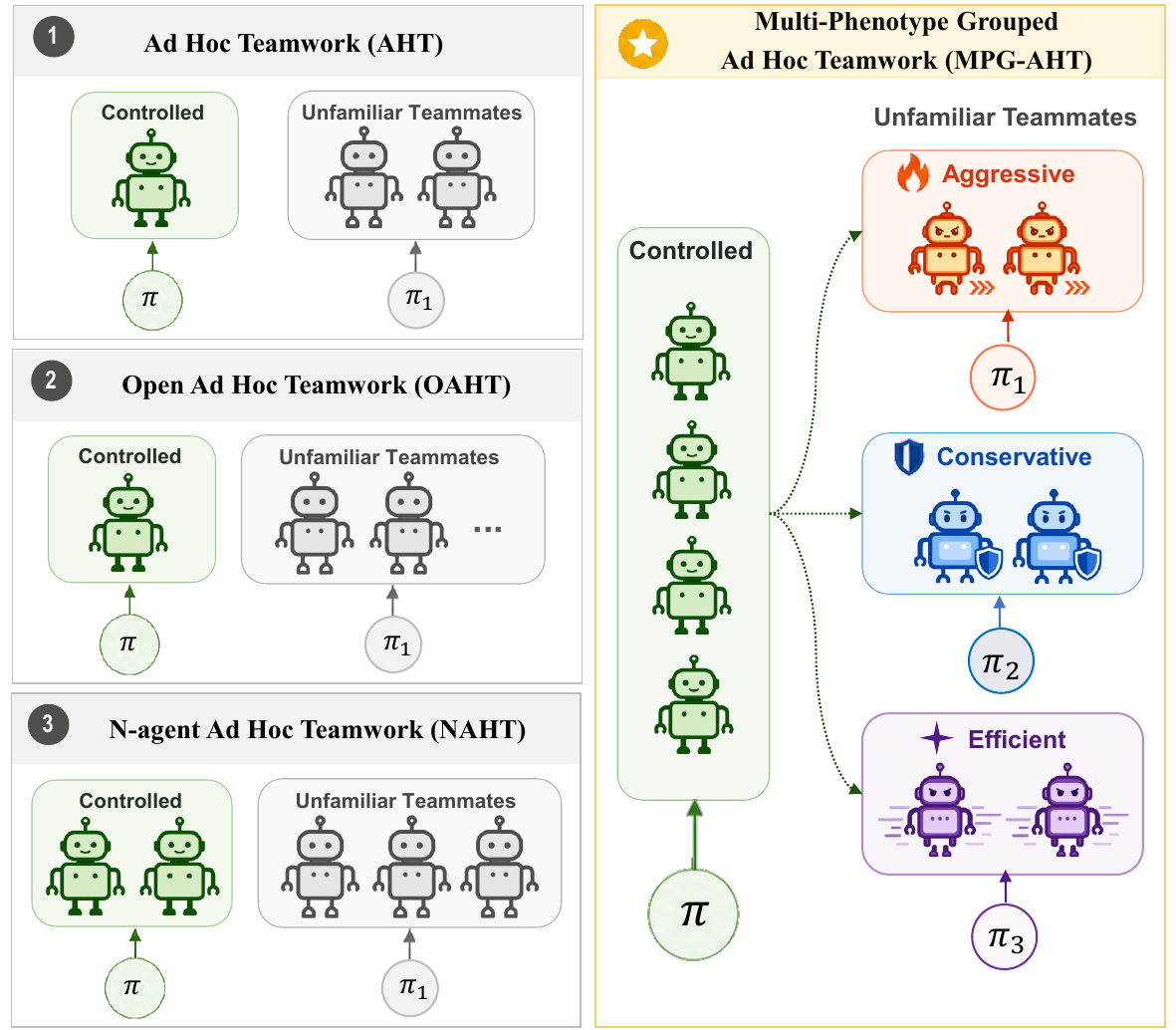}
	\caption{Comparison of AHT paradigms. 
		Prior paradigms (AHT, OAHT, NAHT) consider collaboration with teammate groups exhibiting a single coordination phenotype shaped by the same reward function, 
		whereas MPG-AHT enables collaboration with multiple teammate groups exhibiting diverse coordination phenotypes.}
	\label{fig:MPG-AHT}
\end{figure}

Although AHT and its variants have achieved notable advances, they remain limited by the assumptions about teammate coordination phenotypes. As shown in Figure~\ref{fig:MPG-AHT}, Open AHT (OAHT) focuses on settings where uncontrolled agents may join or leave during training and execution~\cite{GPL,OpenAH_Game}, while $N$-agent AHT (NAHT) considers multiple controlled agents~\cite{poam,poam_2}. 
However, these paradigms still implicitly assume that uncontrolled teammates exhibit a single coordination phenotype shaped by the same reward function, which limits their applicability in realistic open environments.

In many real-world applications, teams are formed from multiple agent groups that have been trained separately and have no prior experience working together. For example, in autonomous surveillance and interception, independently trained groups from different vendors often need to operate in the same mission even though some have learned to approach targets aggressively, while others maintain conservative coordination~\cite{Khan2018CooperativeRobots}. Similar findings also appear in behavioral economics and social decision-making: different populations often develop internally consistent but behaviorally distinct coordination phenotypes due to variations in interaction contexts and incentives~\cite{Peysakhovich2014CooperativePhenotype}. 
As a result, agents often need to collaborate with multiple teammate groups that exhibit distinct coordination phenotypes and are mutually unfamiliar.

To formalize this problem, we introduce \textit{Multi-Phenotype Grouped Ad Hoc Teamwork (MPG-AHT)}, in which controlled agents collaborate with multiple groups of uncontrolled teammates. Each group is associated with a coordination phenotype and has learned to coordinate internally. The central challenge in MPG-AHT is that controlled agents must not only adapt to unfamiliar teammates, but also facilitate collaboration across teammate groups that are themselves mutually unfamiliar. This requires controlled agents to identify coordination phenotypes.

Toward this goal, we propose \textit{Phenotype-Aware Contrastive Team Representation (PACT)} to distinguish coordination phenotypes and capture inter-agent interactions in MPG-AHT.
Inspired by cognitive science studies of social group structure~\cite{Hawkins2021RespectCode}, PACT models individual teammate behaviors from local observations. Meanwhile, it uses phenotype-aware contrastive learning to distinguish different coordination phenotypes while aligning agents that exhibit similar coordination phenotypes. Finally, PACT reasons over relations among agents to form team representations. Subsequently, training policies and value functions conditioned on these team representations enables controlled agents to collaborate more fluently across diverse and mutually unfamiliar teammate groups.

This paper makes three key contributions:
\begin{itemize}
\item We introduce MPG-AHT, a new AHT paradigm that models open environments with multiple teammate groups exhibiting diverse coordination phenotypes and mutual unfamiliarity.
\item We propose PACT, a new training framework for MPG-AHT. PACT is able to learn team representations with agent modeling, phenotype-aware learning, and relational reasoning so as to achieve fluent collaboration across diverse teammate groups.
\item We conduct extensive experiments to evaluate PACT against baseline methods. PACT improves performance, generalization, and robustness, and achieves mean gains of 21.0\% in out-of-distribution performance and 36.5\% in sample efficiency (attached in appendix).
\end{itemize}

\section{Related Work}
\label{sec:related_work}
\noindent\textbf{Multi-agent reinforcement learning.}
MARL studies how multiple agents learn to coordinate in complex collaborative tasks.
Most existing MARL methods assume a fixed set of agents that are jointly trained and fully controllable.
Existing MARL methods fall into two categories.
The first is value-decomposition methods \cite{vdn,qmix,MARL_Qtran,Liu2023NA2QNA}, which decompose the global Q-function into local Q-functions. 
During centralized training, the global Q-function is optimized to guide the learning of local Q-functions, enabling decentralized decision-making during execution. 
The second is actor--critic methods \cite{mpe,mappo,Wang2022MoreCT}, where a centralized critic evaluates the global state and provides gradient signals to update a decentralized actor. 
However, these methods typically assume that \textit{all} agent networks (e.g., local Q-functions or actor) are controllable and updated during training, making it difficult to support ad hoc collaboration with pre-trained, uncontrolled teammate groups learned under different algorithms or objectives.

\paragraph{Ad Hoc Teamwork.}
AHT \cite{FirstAht} studies how a controlled agent can collaborate with unfamiliar teammates \cite{Survey}. 
Existing work has explored different ways to address this challenge, such as training controlled agents to adapt to unfamiliar teammates by learning policies conditioned on embeddings representing teammate behaviors \cite{AgentModeling,Rahman2022AGL}, or generating a diverse set of training teammates to improve the generalization of controlled agents by encouraging interaction with diverse teammates \cite{ROTATE}.

To capture more complex teamwork scenarios, several AHT variants have been proposed. OAHT extends the standard setting to open teams where the number of teammates changes over time \cite{GPL,Rahman2022AGL}. 
Existing methods use graph-based policy learning to model inter-agent interactions and generalize across varying team compositions \cite{OpenAH_Game}.
NAHT \cite{poam,poam_2} further generalizes AHT by allowing multiple controlled agents to collaborate with multiple uncontrolled teammates. 
Policy Optimization with Agent Modelling (POAM) \cite{poam} tackles this setting by learning representations that explicitly model the behaviors of uncontrolled teammates, while Shapley Machine \cite{Wang2025ShapleyMA} leverages Shapley value-based credit assignment to estimate each agent’s contribution.
However, these extensions typically assume that uncontrolled teammates follow a single shared coordination phenotype. 
Therefore, when controlled agents must collaborate with multiple teammate groups exhibiting different coordination phenotypes, these methods are insufficient to address such scenarios.

\section{MPG-AHT Problem Formulation}
\label{sec:problem_formulation}
We formulate the MPG-AHT setting based on a Decentralized Partially Observable Markov Decision Process (Dec-POMDP) \cite{MDP}, a standard framework for collaborative multi-agent decision-making under partial observability.
A Dec-POMDP is defined by the tuple
$\langle \mathcal{N}, \mathcal{S}, \mathcal{A}, \mathcal{O}, T, R, \Omega, \gamma \rangle$,
where $\mathcal{N}$ denotes the set of $n$ agents and $\mathcal{S}$ the state space.
$\mathcal{A} = \prod_{i \in \mathcal{N}} \mathcal{A}_i$ and
$\mathcal{O} = \prod_{i \in \mathcal{N}} \mathcal{O}_i$ denote the joint action and observation spaces.
$T(s' \mid s,\mathbf{a})$ is the transition function, $r_t = R(s,\mathbf{a})$ is the shared team reward, $\Omega(o \mid s,\mathbf{a})$ is the observation function, and $\gamma$ is the discount factor.

MPG-AHT considers a setting where a set of controlled agents collaborates with unfamiliar teammate groups exhibiting diverse coordination phenotypes.
Let $\mathcal{C} \subseteq \mathcal{N}$ denote the set of controlled agents whose policies are optimized during training, and let $\mathcal{U} = \{U_1,\dots,U_L\}$ denote $L$ pre-trained uncontrolled teammate groups, where $\mathcal{C} \cap U_\ell = \emptyset$, $U_\ell \cap U_{\ell'} = \emptyset$ for $\ell \neq \ell'$.
Agents within each group are trained together and are mutually familiar, whereas different groups are trained independently and are mutually unfamiliar.

For each episode, a stochastic \textit{procedure for sampling teams} $X$ generates the team composition by combining controlled agents with uncontrolled teammates instantiated from sampled groups in $\mathcal{U}$.
Specifically, $X$ first samples the number $m$ of uncontrolled teammate groups and the group sizes $(c_1,\dots,c_m,c_{\mathcal{C}})$, where $c_k$ is the number of uncontrolled agents in the $k$-th uncontrolled group, $c_{\mathcal{C}}$ is the number of controlled agents, such that $\sum_{k=1}^{m} c_k + c_{\mathcal{C}} = n$.
For each uncontrolled group $k$, $X$ then selects a group $U_{\ell_k}$ from $\{U_1,\dots,U_L\}$ to determine its coordination phenotype, and draws $c_k$ agents.
The remaining $c_{\mathcal{C}}$ agents are instantiated from the controlled-agent group $\mathcal{C}$.
Together, these agents form the $n$-agent team for the episode and are randomly assigned to the environment slots before deployment.
The detailed sampling procedure is provided in Appendix~\ref{app:sampling}.

Formally, the objective of the MPG-AHT problem is to learn parameters $\theta$ for the controlled-agent policy $C(\theta)$ so as to maximize the expected return under $X$:
\begin{equation}
	\label{eq:obj} \max_{\theta} \left( \mathbb{E}_{\bm{\pi}^{(n)} \sim X(U_1,\dots,U_L, C(\theta))} \left[\sum_{t=0}^{H} \gamma^t r_t \right] \right),
\end{equation}
where $H$ denotes the episode horizon.

The challenges are (i) adapting to unfamiliar groups with different phenotypes (\textit{multi-phenotype generalization}), (ii) collaborating with mutually unfamiliar groups (\textit{group generalization}), and (iii) handling varying numbers of uncontrolled teammates across episodes (\textit{composition variability}).

\paragraph{A motivating example game.}
The following motivating game illustrates why optimizing controlled-agent policies under traditional AHT does not guarantee optimality under the MPG-AHT paradigm.
Consider a two-stage cooperative game with one controlled agent and two uncontrolled teammates with phenotypes $p_1,p_2\in\mathcal{P}=\{A,B,C\}$.
In the first stage, the controlled agent infers these phenotypes from local interaction history.
In the second stage, it chooses $a\in\{0,1\}$, and the reward is $r=1$ if $a=\Gamma(p_1,p_2)$ and $r=0$ otherwise, where $\Gamma$ is symmetric:
\[
\begin{aligned}
\Gamma(A,A)&=\Gamma(B,B)=\Gamma(C,C)=1,\\
\Gamma(A,B)&=\Gamma(B,C)=1,\\
\Gamma(A,C)&=0.
\end{aligned}
\]

Under the traditional AHT paradigm, suppose that both uncontrolled teammates are sampled from the same phenotype group, such that $p_1=p_2$. Since $\Gamma(p,p)=1$ for every $p\in\mathcal{P}$, the policy $\pi_{\mathrm{AHT}}$ that always chooses $a=1$ is optimal, with $J_{\mathrm{AHT}}(\pi_{\mathrm{AHT}})=1$.
However, under MPG-AHT with two uncontrolled teammate groups ($m=2$), the group phenotype pairs $(A,B)$, $(A,C)$, and $(B,C)$ occur with equal probability across episodes.
The policy $\pi_{\mathrm{AHT}}$ only succeeds on the first and third pairs, whereas the optimal MPG-AHT policy chooses $a=\Gamma(p_1,p_2)$ and succeeds on all three:
\[
J_{\mathrm{MPG}}(\pi_{\mathrm{AHT}})=\frac{2}{3}<1=J_{\mathrm{MPG}}^\star.
\]
Thus, optimality under the traditional AHT paradigm does not guarantee optimality under the MPG-AHT paradigm.
The example also indicates that effective MPG-AHT policies should distinguish coordination phenotypes to make adaptive decisions.

\begin{figure*}[t]
	\centering
	\includegraphics[width=1.0\textwidth]{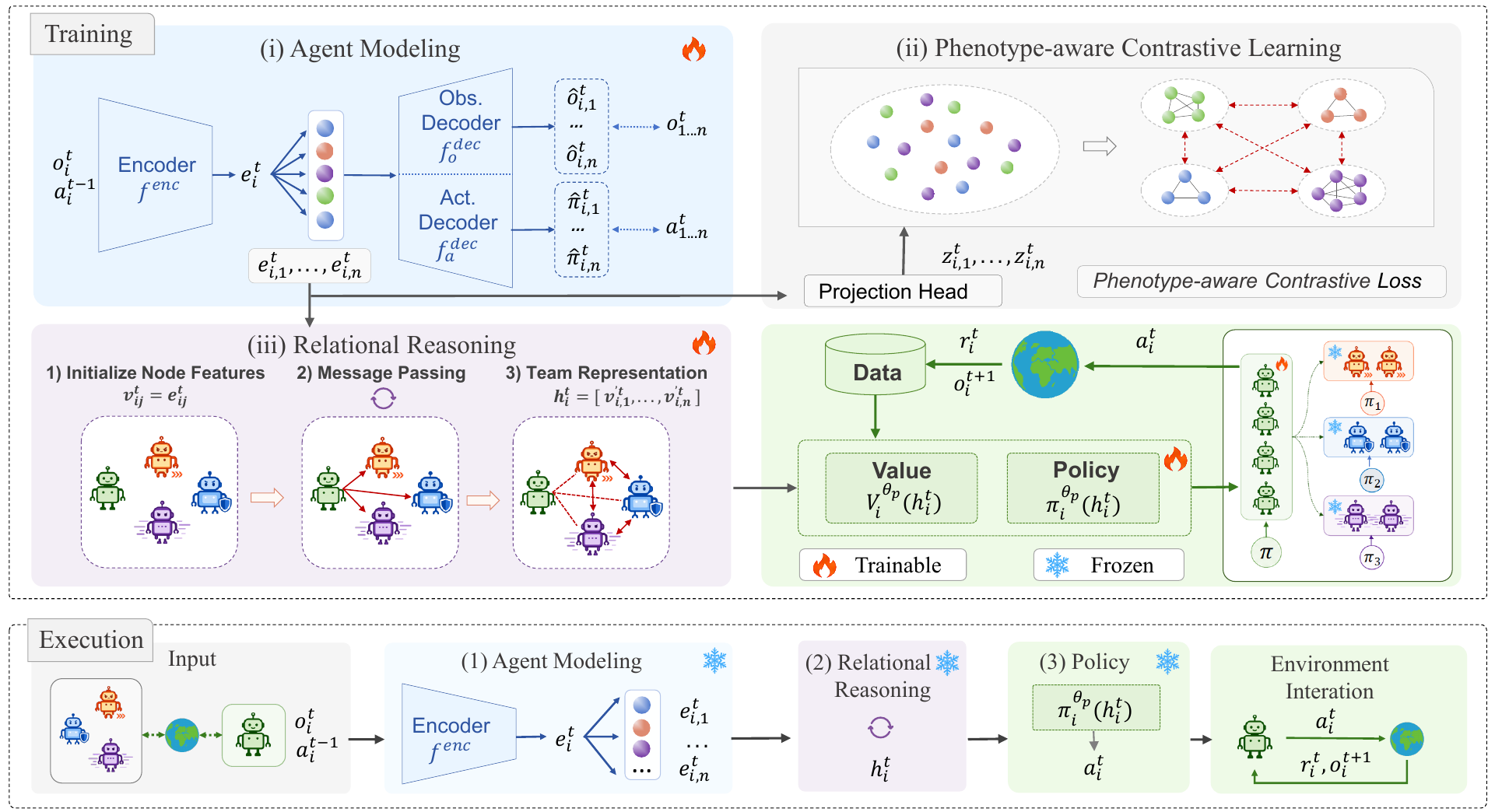}
			\caption{
				Overview of the PACT framework.
				During centralized training, PACT combines agent modeling, phenotype-aware contrastive learning, and relational reasoning to derive team representations $h_i^t$.
				During decentralized execution, each controlled agent infers $h_i^t$ from its local history.
			}
	\label{fig:star}
\end{figure*}

\section{PACT for MPG-AHT}

Designed to solve the challenges of MPG-AHT, we introduce PACT, a new training framework.
PACT is motivated by cognitive science studies showing that people learn how individual partners communicate through interaction and expect the resulting conventions to extend to members of the same social group, but not to members of different social groups~\cite{Hawkins2021RespectCode}.
As illustrated in Figure~\ref{fig:star}, PACT first uses agent modeling to infer the behavioral features of each agent from observed interactions. Phenotype-aware contrastive learning then brings behavioral features from the same coordination phenotype closer while separating those across phenotype groups. Relational reasoning further models how these individual behaviors interact within a team to derive team representations. Training policies and value functions conditioned on these team representations enables adaptation to multi-phenotype teammate groups.

\subsection{Agent Modeling}

In MPG-AHT, controlled agents must collaborate with unfamiliar teammates whose policies are inaccessible. Therefore, PACT employs an encoder-decoder agent modeling module to infer behavioral features from each controlled agent's local interaction history.

For each controlled agent $i$, we define the local interaction history as 
$\xi_i^t = \{o_i^k, a_i^{k-1}\}_{k=1}^t$, 
where $o_i^k$ and $a_i^{k-1}$ denote agent $i$'s local observation and previous action at timestep $k$. 
The encoder $f_{\theta^e}^{\mathrm{enc}}$ maps this local history to a feature $e_i^t = f_{\theta^e}^{\mathrm{enc}}(\xi_i^t)$, from which controlled agent $i$ infers a latent feature for each agent $j$:
\[
e_{i,j}^t = g_{\phi}(e_i^t, \mathrm{id}_j),
\]
where $\mathrm{id}_j$ denotes the one hot slot index of agent $j$.

To ensure that the inferred features contain sufficient information about agent behaviors, two decoder networks are introduced to reconstruct the current observation and predict the agent's policy over actions. For each target agent $j$, the observation decoder $f_{\theta^o}^{\mathrm{dec}}$ reconstructs the current observation 
$\hat{o}_{i,j}^t = f_{\theta^o}^{\mathrm{dec}}(e_{i,j}^t)$, 
while the action decoder $f_{\theta^a}^{\mathrm{dec}}$ predicts the agent's action distribution
$\hat{\pi}_{i,j}^t = f_{\theta^a}^{\mathrm{dec}}(e_{i,j}^t)$.

The agent modeling loss is defined as:
\begin{equation}
	L_{\mathrm{AE}} =
		\sum_{j=1}^{n}
		\| \hat{o}_{i,j}^t - o_j^t \|^2
		-
		\sum_{j=1}^{n}\log \hat{\pi}_{i,j}^t(a_j^t).
\end{equation}

\subsection{Phenotype-aware Contrastive Learning}

Although the agent modeling module infers behavioral features of unfamiliar teammates from observations, agents with different coordination phenotypes are often still mapped to nearby locations in the feature space, making it difficult to distinguish their behavioral preferences.
To better capture coordination phenotypes, PACT applies a contrastive learning module, encouraging the learned features to be discriminative across phenotypes.

To make the contrastive comparison depend mainly on feature directions rather than magnitudes, we compute the contrastive loss in a normalized projection space~\cite{DBLP:conf/nips/KhoslaTWSTIMLK20}.
Specifically, each feature is mapped as:
\[
z_{i,j}^t = \frac{g_{\psi}(e_{i,j}^t)}{\|g_{\psi}(e_{i,j}^t)\|_2},
\]
where $g_{\psi}$ is a learnable projection head.
Each projected feature $z_{i,j}^t$ is associated with a phenotype label $y_{i,j}^t$, defined by its reward-shaping configuration and used only during training.

Within each minibatch, we collect all projected features and write them as $\{(z_q, y_q)\}_{q=1}^{N}$, where $z_q$ is one projected feature and $y_q$ is its phenotype label.
For the projected feature $z_q$ used as an anchor, we define the positive set $\mathcal{P}(q)$ as all other projected features with the same phenotype label:
\[
\mathcal{P}(q)=\{r\in\{1,\ldots,N\}\setminus\{q\}\mid y_r=y_q\},
\]
and $\mathcal{A}=\{q\mid |\mathcal{P}(q)|>0\}$ is the set of anchors that have at least one positive sample in the minibatch.

Let $\kappa_{q,s}=\exp(\mathrm{sim}(z_q,z_s)/\tau)$.
We introduce a phenotype-aware contrastive objective as:
\begin{equation}
\label{eq:contrastive}
\begin{aligned}
L_{\mathrm{con}}
=&-\frac{1}{|\mathcal{A}|}
\sum_{q\in\mathcal{A}}
\frac{1}{|\mathcal{P}(q)|}
\sum_{r\in\mathcal{P}(q)}
\log
\frac{\kappa_{q,r}}{\sum_{s\neq q}\kappa_{q,s}},
\end{aligned}
\end{equation}
where $\mathrm{sim}(\cdot,\cdot)$ is cosine similarity and $\tau$ is a temperature parameter.
Additional analysis of this objective is provided in Appendix~\ref{app:phenotype_theory}.

\subsection{Relational Reasoning}

To facilitate collaboration among unfamiliar teammates, PACT employs a relational reasoning module that models the multi-phenotype teams and captures their interactions using a graph structure.

Formally, for controlled agent $i$, the graph is represented as $(u_i, V_i, E_i)$, where $u_i$ denotes the global attribute, $V_i=\{v_{i,j}\}_{j=1}^{n}$ the node attributes, and $E_i=\{(m_{i,j,k},r_{i,j,k},s_{i,j,k})\}$ the edge attributes $m_{i,j,k}$ with receiver $r_{i,j,k}$ and sender $s_{i,j,k}$ indices. 
Each node is initialized with the latent feature $e_{i,j}^t$. 
Edge and global attributes are initialized as learnable vectors.

The relational reasoning module updates edge, node, and global attributes through graph network message passing and aggregation operators~\cite{GraphNetworks}:
\begin{align}
	\label{eq:rr_edge}
	m_{i,j,k}' &= \phi^e(m_{i,j,k}, v_{i,r_{i,j,k}}, v_{i,s_{i,j,k}}, u_i), \\
	\label{eq:rr_node}
	v_{i,j}^{\prime t} &= \phi^v(\rho^{m\to v}(E'_{i,j}), v_{i,j}^t, u_i), \\
	\label{eq:rr_global}
	u_i' &= \phi^u(\rho^{m\to u}(E_i'), \rho^{v\to u}(V_i'), u_i),
\end{align}
where $\phi$ represents learnable update functions and $\rho$ represents aggregation operators.

After one iteration of message passing, the relational reasoning module captures interactions between controlled and uncontrolled agents. To model interactions of higher order, the module then performs another iteration of message passing. This iterative process produces updated node features $\{v_{i,j}^{\prime t}\}_{j=1}^{n}$ from the perspective of controlled agent $i$. 
We concatenate all updated node features to form the team representation used by controlled agent $i$:
\[
h_i^t = [v_{i,1}^{\prime t}, v_{i,2}^{\prime t}, \ldots, v_{i,n}^{\prime t}].
\]
This representation retains the relational features of all modeled agents, allowing the policy and value functions to condition on a complete team-level view inferred from the controlled agent $i$.

\subsection{Policy and Value Networks}
In PACT, each controlled agent conditions its policy and value functions on both the team representation $h_i^t$ defined above and the local interaction history $\xi_i^t$. Since the number of controlled agents in MPG-AHT varies and the learned representation already summarizes interactions with unfamiliar teammates, we adopt IPPO~\cite{Witt2020IsIL} with shared parameters across controlled agents and parameterize them as:
\[
\pi_i^{\theta^p}(a_i^t \mid h_i^t, \xi_i^t)
\quad \text{and} \quad
V_i^{\theta^c}(h_i^t, \xi_i^t).
\]

The overall PACT objective integrates all components:
$L_{\mathrm{policy}}$ and $L_{\mathrm{value}}$ are the standard IPPO policy optimization loss and value regression loss, respectively.
\begin{equation}
	L_{\mathrm{PACT}}
	=
	L_{\mathrm{policy}}
	+
	\lambda_v L_{\mathrm{value}}
	+
	\lambda_{\mathrm{AE}} L_{\mathrm{AE}}
	+
	\lambda_{\mathrm{con}} L_{\mathrm{con}}.
\end{equation}

\section{Experiments}
\begin{figure*}[t]
	\centering
	\includegraphics[width=1.0\textwidth]{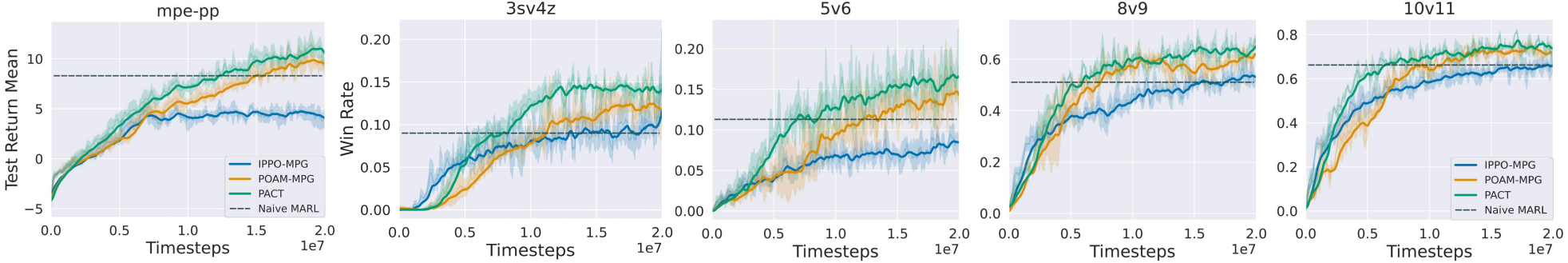}
	\caption{
		Test return mean and win rate on mpe-pp and four scenarios of SMAC}
	\label{fig:main_result1}
\end{figure*}

In this section, we evaluate PACT against representative MARL and AHT baselines, and examine generalization to OOD evaluation and robustness to phenotype gaps and varying team compositions.
We also analyze the contribution of PACT through ablation studies and representation visualizations, and evaluate its transferability by integrating PACT with another AHT baseline.
Additional experimental details and supplementary results, including hyperparameter sensitivity, training efficiency, and battle outcome analysis, are provided in Appendix~\ref{app:appendix_experiments} and Appendix~\ref{app:supplementary_experimental_results}.
\subsection{Experimental Design}
\label{sec:ExpDesign}

\newcommand{\tabci}[2]{$#1{\pm}#2$}
\newcommand{\besttabci}[2]{$\bm{#1{\pm}#2}$}
\newcommand{\secondtabci}[2]{\underline{\tabci{#1}{#2}}}
\newcommand{\tabciunknown}[1]{$#1{\pm}\text{--}$}
\newcommand{\besttabciunknown}[1]{$\bm{#1{\pm}\text{--}}$}
\newcommand{\secondtabciunknown}[1]{\underline{\tabciunknown{#1}}}

\begin{table*}[h]
	\centering
	\small
	\begin{tabular*}{\textwidth}{@{\extracolsep{\fill}}lcccccccc@{}}
		\toprule
		\multirow{2}{*}{Map}
		& \multicolumn{5}{c}{\textit{Naive MARL Baselines}}
		& \multicolumn{3}{c}{\textit{MPG-AHT Variants}} \\
		\cmidrule(lr){2-6}\cmidrule(lr){7-9}
		& IPPO & MAPPO & IQL & VDN & QMIX & IPPO-MPG & POAM-MPG & PACT\\
		\midrule
		mpe-pp & \tabci{1.61}{0.86} & \tabci{1.80}{0.91} & \tabci{1.56}{0.72} & \tabci{2.12}{0.37} & \tabci{1.96}{0.55} & \tabci{6.50}{2.63} & \secondtabci{10.10}{3.21} & \besttabci{12.80}{3.85} \\
		3sv4z & \tabci{2.26}{3.39} & \tabci{3.59}{3.60} & \tabci{1.47}{2.01} & \tabci{3.07}{2.14} & \tabci{3.34}{4.85} & \tabci{9.15}{4.37} & \secondtabci{9.89}{3.72} & \besttabci{13.21}{4.35} \\
		5v6 & \tabci{2.65}{1.04} & \tabci{3.56}{2.10} & \tabci{1.26}{0.73} & \tabci{2.78}{1.19} & \tabci{3.56}{1.27} & \secondtabci{8.31}{5.66} & \tabci{7.87}{4.93} & \besttabci{10.84}{5.72} \\
		8v9 & \tabci{38.85}{5.18} & \tabci{39.05}{4.77} & \tabci{22.85}{4.74} & \tabci{45.62}{7.03} & \tabci{45.86}{6.88} & \tabci{44.27}{11.30} & \secondtabci{47.68}{12.85} & \besttabci{51.82}{10.66} \\
		10v11 & \tabci{60.60}{3.63} & \tabci{62.78}{3.45} & \tabci{60.14}{4.52} & \tabci{50.88}{6.15} & \secondtabci{63.30}{4.72} & \tabci{49.72}{11.87} & \tabci{60.25}{8.91} & \besttabci{66.84}{10.12} \\
		\bottomrule
	\end{tabular*}
	\caption{
		Performance across environments under OOD teammate settings. 
		\texttt{mpe-pp} reports test return, while SMAC tasks report win rate (\%) in the form mean $\pm$ 95\% CI.
	}
	\label{tab:maht_win}
\end{table*}

\noindent\textbf{Experimental Environments.} We evaluate PACT on two widely used benchmarks in MARL: Multi-Agent Particle~\cite{mpe} Predator--Prey (\texttt{mpe-pp}) and four StarCraft Multi-Agent Challenge (SMAC)~\cite{SMAC} tasks: \texttt{3sv4z}, \texttt{5v6}, \texttt{8v9}, and \texttt{10v11}.
In \texttt{mpe-pp}, three predators cooperate to catch a prey agent controlled by the server. 
In SMAC, allied agents must defeat enemies controlled by the server.
More details about environments are provided in Appendix~\ref{app:env}.

\noindent\textbf{Uncontrolled Teammate Groups.} Uncontrolled teammates are organized into several coordination phenotypes.
Different phenotypes are induced by distinct reward shaping configurations during teammate training, and represent different coordination tendencies such as more aggressive or more risk-sensitive behavior.
To improve the realism and representativeness of the teammate pool, we train the policies used to instantiate each phenotype with different MARL algorithms.
Specifically, we use five representative MARL algorithms: IPPO, MAPPO~\cite{mappo}, IQL~\cite{iql}, VDN~\cite{vdn}, and QMIX~\cite{qmix}.
The detailed reward configurations are provided in Appendix~\ref{app:phenotype_teammates}.

The uncontrolled teammate groups are further partitioned into two disjoint sets: a training set $U_{\mathrm{train}}$ and a test set $U_{\mathrm{test}}$. 
The set $U_{\mathrm{train}}$ is used to train the controlled agents and to evaluate their training-time coordination performance with unfamiliar teammates, following the standard setting in AHT. We use $U_{\mathrm{test}}$ to assess the generalization ability of trained controlled agents when collaborating ad hoc with out-of-distribution (OOD) teammates.

\noindent\textbf{MPG-AHT Training Setting.} In our experiments, controlled agents are homogeneous, share parameters, and act based only on local observations. 
During both training and evaluation, controlled agents collaborate with teammate groups sampled by the team-sampling procedure $X$ described in Section~\ref{sec:problem_formulation}.

In the main experiments, we focus on an MPG-AHT setting where controlled agents collaborate with two uncontrolled teammate groups that exhibit different coordination phenotypes and are unfamiliar with one another.
All experiments are conducted with five independent random seeds. For SMAC tasks, we report the mean win rate, while for \texttt{mpe-pp} we report the mean test return. Shaded regions indicate 95\% confidence intervals.

\noindent\textbf{Baselines.} We compare PACT with representative MARL and AHT baselines.
Unless otherwise stated, all methods are trained under the same MPG-AHT training setting.
\begin{itemize}
	
	\item Naive MARL. 
	We consider representative MARL algorithms including IPPO, MAPPO, IQL, VDN, and QMIX. 
	These methods are trained in self-play, where agents repeatedly coordinate with fixed teammates under a single learning algorithm, and directly evaluated in the MPG-AHT environment to assess how conventional MARL performs under multi-phenotype collaboration.
	For OOD evaluation, each trained naive MARL policy is tested with unseen teammate groups in $U_{\mathrm{test}}$ generated by other naive MARL algorithms.
	
	\item IPPO-MPG:
	Since IPPO treats other agents as part of the environment, it is naturally applicable to AHT scenarios. 
	We therefore train IPPO directly under the MPG-AHT setting, providing a naive baseline to evaluate the benefit of team representation.
	
	\item POAM-MPG:
	POAM is a representative AHT method that learns behavioral representations of teammates through agent modeling, but does not explicitly model structural relationships among multiple teammate groups or the diversity of coordination phenotypes.
\end{itemize}
We further study Shapley Machine as an orthogonal value learning framework based on Shapley-value credit assignment in Section~\ref{sec:shapley_integration}.

\subsection{Main Results}
\label{sec:exp_core_results}
Figure~\ref{fig:main_result1} shows the testing curves of the controlled agents under the MPG-AHT setting. 
The horizontal dashed lines indicate the test win rate achieved by the best naive MARL baseline. 
Across all tasks, PACT consistently learns faster than the baselines.
POAM-MPG eventually reaches competitive performance in some cases but converges more slowly, as its agent embeddings capture individual behaviors without modeling structured relations among agents. 
IPPO-MPG improves faster early on, since treating other agents as part of the environment simplifies learning. However, its performance soon plateaus and is later surpassed by both PACT and POAM-MPG, because without explicit teammate modeling, changing teammate behaviors across phenotype mixtures make the learning target unstable. On \texttt{3sv4z} and \texttt{5v6}, PACT maintains the strongest learning curve throughout training, while on \texttt{8v9} and \texttt{10v11} it converges faster, especially in the early and middle stages.

\paragraph{Out-of-Distribution Generalization.}
Beyond training-time coordination, we test all methods under an OOD setting related to zero-shot coordination (ZSC), where controlled agents interact with teammate groups whose phenotypes are unseen during training.
These test-time groups are generated using different reward shaping configurations, as described in Appendix~\ref{app:phenotype_teammates}. Table~\ref{tab:maht_win} summarizes the OOD results across environments, where the best and second-best results are highlighted in \textbf{bold} and \underline{underlined}, respectively.
PACT achieves the highest mean OOD performance on all evaluated tasks, suggesting that its phenotype-aware representation helps transfer coordination knowledge to unseen coordination behaviors.
On average, PACT achieves a 21.0\% OOD performance gain, computed by comparing it with the best baseline and averaging these gains across the five tasks.
Naive MARL and IPPO-MPG do not explicitly condition their policies on teammate representations, while POAM-MPG models teammate behavior without explicitly separating phenotype structure.
Together, these results show that PACT improves both training-time coordination and generalization to held-out phenotypes.
Detailed analyses of training efficiency and final battle outcomes are provided in Appendix~\ref{app:training_battle_analysis}.
\begin{figure*}[t]
	\centering
	\includegraphics[width=1.0\textwidth]{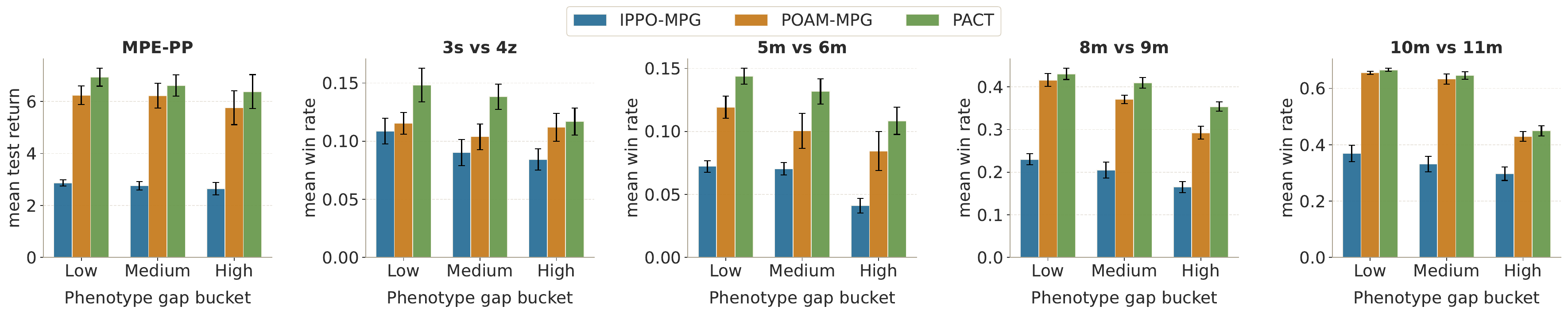}
	\caption{
		Performance under low, medium, and high phenotype gap buckets across all tasks.}
	\label{fig:phenotype_gap_robustness}
\end{figure*}

\begin{figure*}[h]
	\centering
	\includegraphics[width=\textwidth]{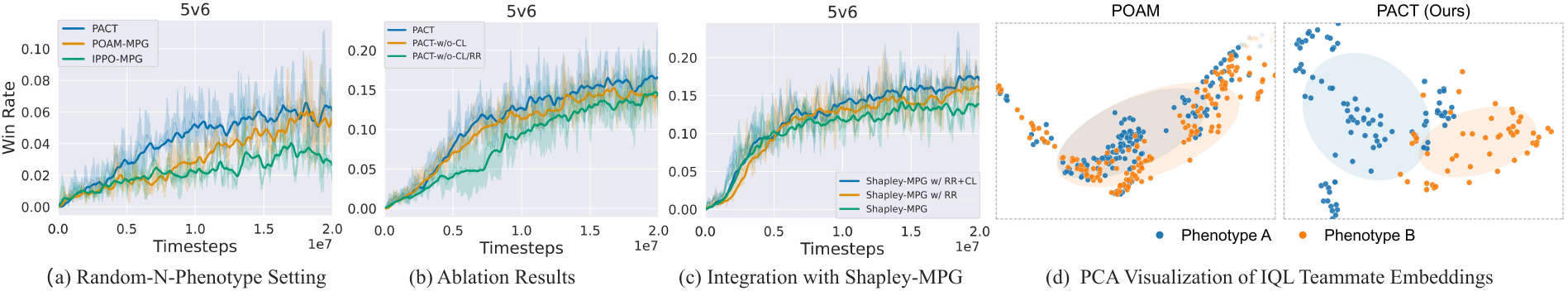}
	\caption{
		Additional analysis results.
		(a) Random-N-Phenotype results on the SMAC \texttt{5v6} map.
		(b) Ablation results on the SMAC \texttt{5v6} map.
		(c) Integration results with Shapley-MPG on the SMAC \texttt{5v6} map.
		(d) PCA visualization of POAM encoder features and PACT projection features for IQL teammates with different phenotypes.}
	\label{fig:generation}
\end{figure*}

\subsection{Robustness Analysis}
\label{sec:robustness_analysis}
We further analyze robustness under different phenotype gap and team composition.

\paragraph{Phenotype Gap Robustness.}
\label{sec:phenotype_gap_robustness}
We further test how PACT performs when the coordination phenotypes of teammate groups are either similar or substantially different.
To quantify this difference, we define the phenotype gap as the behavioral distance between uncontrolled teammate groups with different coordination phenotypes, computed from observed rollout behavior.
The detailed behavioral distance definition and bucket construction are provided in Appendix~\ref{app:phenotype_teammates}.
As shown in Figure~\ref{fig:phenotype_gap_robustness}, larger phenotype gaps generally make collaboration harder, especially for IPPO-MPG and POAM-MPG.
This is expected because groups with larger behavioral distances are more likely to follow incompatible engagement timing, spacing, or risk preferences.
PACT generally achieves higher average performance and keeps its advantage when the phenotype gap is medium or large, because contrastive learning keeps phenotype information clear and relational reasoning helps coordinate different groups.

\paragraph{Random-N-Phenotype Settings.}
\label{sec:Random-N-Phenotype}
To evaluate robustness under varying team compositions and a more general MPG-AHT setting, we conduct a Random-N-Phenotype experiment on the \texttt{5v6} task in SMAC, where Random-N means that the number $N$ of uncontrolled teammate groups varies across episodes. 
Specifically, $N$ is randomly sampled from $\{2,3,4\}$ for each episode, as the task contains at most five collaborating agents. 
Other experiment details follow the experimental setup described in Section~\ref{sec:ExpDesign}.

As shown in Figure~\ref{fig:generation}(a), PACT consistently achieves the strongest performance during training. 
Due to the increased randomness in team composition and the complexity of ad hoc coordination, all methods experience some performance degradation compared with the fixed two-group setting. 
Nevertheless, PACT maintains clear advantages, while POAM-MPG and IPPO-MPG struggle to adapt to the changing team compositions.

\subsection{Component Analysis}
\label{sec:component_analysis}

\paragraph{Ablation Study.}
\label{sec:ablation_study}
We analyze the contributions of PACT's representation modules.
We conduct an incremental ablation study by progressively adding PACT's representation modules to a base variant without phenotype-aware contrastive learning (CL) or relational reasoning (RR).
Specifically, PACT w/o CL/RR removes both modules, PACT w/o CL adds RR back, and full PACT further adds CL. 
As shown in Figure~\ref{fig:generation}(b), PACT consistently achieves the best performance. 
Compared with PACT w/o CL/RR, adding RR improves coordination by enabling the policy to reason over interactions among unfamiliar teammate groups.
Further adding CL yields the strongest performance because teammate features from different phenotypes become more separable.

To further examine what the CL module contributes to the learned representation, Figure~\ref{fig:generation}(d) visualizes teammates trained with the same underlying algorithm (IQL) but different coordination phenotypes. 
In POAM, the encoder features of the two phenotypes remain substantially overlapped, suggesting that agent modeling alone does not naturally organize the representation space by phenotype. 
In contrast, the projection features learned by PACT show clearer separation because the contrastive objective explicitly pulls samples from the same phenotype together and pushes different phenotypes apart.
The complete PCA and t-SNE visualizations across teammate algorithms are provided in Appendix~\ref{app:phenotype_visualization}.

\paragraph{Integration with Shapley-MPG.}
\label{sec:shapley_integration}
Finally, we test whether the PACT representation can be integrated into another MPG-AHT training framework.
We use Shapley-MPG as the base framework and compare it with two augmented variants: Shapley-MPG with RR and Shapley-MPG with RR and CL.

As shown in Figure~\ref{fig:generation}(c), Shapley-MPG w/ RR consistently outperforms Shapley-MPG on \texttt{5v6}, while Shapley-MPG w/ RR and CL achieves the best performance overall. 
The results suggest that relational reasoning and phenotype-aware contrastive learning complement Shapley-value credit assignment and further improve learning performance when integrated into other downstream policy learning methods.

\section{Conclusion and Future Work} 

This paper studies a realistic collaboration scenario where agents must collaborate with multiple teammate groups that follow different coordination phenotypes and are mutually unfamiliar. We formalize this paradigm as MPG-AHT. To address this challenge, we propose PACT, a new training framework that integrates agent modeling, phenotype-aware contrastive learning, and relational reasoning to distinguish coordination phenotypes and capture inter-agent interactions. Experiments on SMAC and MPE show that PACT improves performance, generalization, robustness, and sample efficiency across diverse team compositions.

Several directions remain for future work. PACT could be combined with complementary coordination mechanisms such as communication learning \cite{Foerster2016LearningTC,soccer} and population-based training \cite{Jaderberg2017PopulationBT} to further enhance coordination in multi-agent populations.

\clearpage
\bibliography{star_refs}

\clearpage
\appendix
\label{sec:appendix}

\twocolumn[
\begin{@twocolumnfalse}
\begin{center}
{\Large\bfseries Appendix}
\end{center}
\vspace{0.5em}
\noindent{\bfseries Contents}\\
\noindent{\bfseries Appendix section}
\vspace{0.25em}
\hrule
\vspace{0.45em}
\noindent\begin{tabular}{@{}p{0.035\textwidth}p{0.90\textwidth}@{}}
{\bfseries A} & {\bfseries Analysis of the Motivating Game and Contrastive Objective} \\
{\bfseries B} & {\bfseries Experimental Details} \\
& B.1 Experimental Environments \\
& B.2 Hyperparameters \\
& B.3 PACT Implementation Details \\
& B.4 Detailed Procedure for Sampling Teams \\
& B.5 Generation of Uncontrolled Teammate Groups with Different Phenotypes \\
& B.6 Evaluation Procedure \\
& B.7 Computational Infrastructure \\
{\bfseries C} & {\bfseries Supplementary Experimental Results} \\
& C.1 Hyperparameter Sensitivity \\
& C.2 Additional Analysis of Training Efficiency and Battle Outcomes \\
& C.3 Additional Visualization of Phenotype Separation \\
\end{tabular}
\vspace{0.45em}
\hrule
\vspace{1.2em}
\end{@twocolumnfalse}
]

\section{Analysis of the Motivating Game and Contrastive Objective}
\label{app:phenotype_theory}

This section complements the analysis in Section~\ref{sec:problem_formulation} and Section~4. We reuse the notation in Eq.~(\ref{eq:obj}) and Eq.~(\ref{eq:contrastive}) and do not introduce any new training objective.
It addresses two questions: what phenotype information is needed in the motivating game, and how Eq.~(\ref{eq:contrastive}) changes similarities between projected features.

\paragraph{Phenotype information in the motivating game.}Consider a representation used to select the final action in the motivating game.
Suppose it maps the configurations $(A,B)$ and $(A,C)$ to the same value and that no other policy input distinguishes these two configurations.
The policy must then use the same action distribution in both cases.
However, $\Gamma(A,B)=1$ and $\Gamma(A,C)=0$, so the two cases require different actions.
Because the action space is binary, identical action distributions imply
\[
\Pr(a=1\mid A,B)+\Pr(a=0\mid A,C)=1.
\]
The three configurations $(A,B)$, $(A,C)$, and $(B,C)$ occur with equal probability, and the reward for $(B,C)$ is at most $1$.
Combining this upper bound with the identity above gives
\[
J_{\mathrm{MPG}}(\pi)
\leq \frac{1}{3}(1+1)
=\frac{2}{3}
<1
=J_{\mathrm{MPG}}^\star.
\]
Thus a representation intended to support the optimal action cannot merge teammate configurations that require different actions.
In particular, recording only whether the two teammates have the same or different phenotypes is insufficient, because all three evaluated pairs contain different phenotypes while $(A,C)$ has a different compatibility value.

\paragraph{Effect of the phenotype-aware contrastive objective.}For compactness, normalize the comparison terms for each anchor $q$ as
\[
\widetilde{\kappa}_{q,s}
=
\frac{\kappa_{q,s}}{\sum_{r\neq q}\kappa_{q,r}}
\qquad (s\neq q).
\]
The contribution of anchor $q$ and the full objective is written as
\[
\begin{aligned}
L_{\mathrm{con}}^{(q)}
&=
-\frac{1}{|\mathcal{P}(q)|}
\sum_{r\in\mathcal{P}(q)}
\log\widetilde{\kappa}_{q,r},\\
L_{\mathrm{con}}
&=
\frac{1}{|\mathcal{A}|}
\sum_{q\in\mathcal{A}}L_{\mathrm{con}}^{(q)}.
\end{aligned}
\]
Using $\kappa_{q,s}=\exp(\mathrm{sim}(z_q,z_s)/\tau)$, direct differentiation gives
\[
\frac{\partial L_{\mathrm{con}}^{(q)}}
{\partial\,\mathrm{sim}(z_q,z_s)}
=
\begin{cases}
\displaystyle
\frac{1}{\tau}
\left(
\widetilde{\kappa}_{q,s}
-
\frac{1}{|\mathcal{P}(q)|}
\right),
& s\in\mathcal{P}(q),\\[8pt]
\displaystyle
\frac{\widetilde{\kappa}_{q,s}}{\tau},
& s\notin\mathcal{P}(q).
\end{cases}
\]
For a sample with a different phenotype label, this derivative is positive, so gradient descent lowers its similarity to the anchor when the similarity scores are considered individually.
For a positive sample, gradient descent raises its similarity when $\widetilde{\kappa}_{q,s}<1/|\mathcal{P}(q)|$ and lowers it when the normalized weight exceeds this value.
Hence Eq.~(\ref{eq:contrastive}) moves comparison weight away from different-phenotype samples and distributes it across same-phenotype samples. It does not necessarily increase every positive similarity at every update.

\paragraph{Implications and limitations.}The motivating game shows that phenotype information is needed when different teammate combinations require different actions.
The gradient calculation then explains how Eq.~(\ref{eq:contrastive}) encourages the projected features to preserve such information by separating samples with different phenotype labels.
These results explain the role of the phenotype-aware contrastive objective, but they do not prove that the learned representation is globally separated, that optimization converges to a global optimum, or that reducing $L_{\mathrm{con}}$ alone must improve the MPG-AHT return.
They also do not claim that contrastive learning is the only possible way to retain phenotype information.

\FloatBarrier
\section{Experimental Details}
\label{app:appendix_experiments}

\subsection{Experimental Environments}
\label{app:env}

\paragraph{MPE Predator--Prey.}We evaluate our method in the Predator--Prey task from the Multi-Agent Particle Environment (MPE) \cite{mpe}, a lightweight platform for studying multi-agent interaction in continuous spaces. 
In this task, three predator agents must collaborate to capture a prey agent that follows a fixed escape policy.

\begin{figure}[h]
	\centering
	\includegraphics[width=0.5\columnwidth]{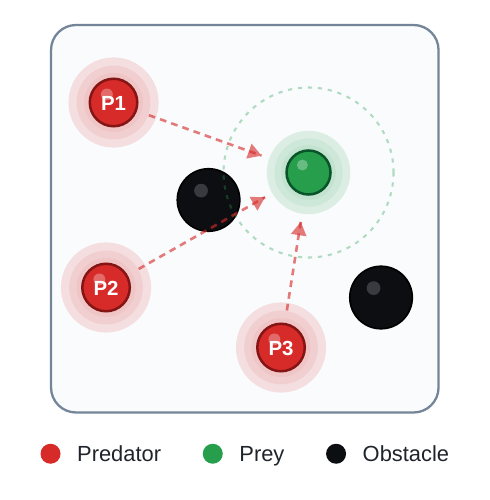}
	\caption{
		Predator--Prey scenario in the Multi-Agent Particle Environment.}
	\label{fig:mpe}
\end{figure}

All agents move in a plane with obstacles. 
As illustrated in Fig.~\ref{fig:mpe}, the red circles denote predator agents, the green circle represents the prey, and the black circles indicate obstacles.

Each predator observes its own state and the relative positions and velocities of other agents. 
The action space consists of four movement directions and a no operation action. 
A collaborative reward is provided when at least two predators simultaneously collide with the prey, with additional shaping rewards encouraging predators to approach the prey. 
Episodes terminate when the prey is captured or when a time limit is reached.

\paragraph{SMAC.}We also evaluate our method on several scenarios from the StarCraft Multi-Agent Challenge (SMAC) benchmark \cite{SMAC}, a widely used collaborative multi-agent reinforcement learning environment based on StarCraft II. 
In SMAC, a team of allied agents must coordinate to defeat enemy units controlled by the game engine under partial observability.

Each allied unit is treated as an independent agent and selects actions based on its local observation during execution. 
Although SMAC provides a centralized global state containing information about all units (e.g., positions, health, and cooldowns), the reported PACT runs do not feed this global state to the critic. 
Instead, PACT uses decentralized actor and value inputs derived from each controlled agent's local history, while centralized training signals are provided through value learning and the auxiliary agent modeling and contrastive objectives.

We consider four SMAC scenarios: \texttt{3s\_vs\_4z} (\texttt{3sv4z}), \texttt{5m\_vs\_6m} (\texttt{5v6}), \texttt{8m\_vs\_9m} (\texttt{8v9}), and \texttt{10m\_vs\_11m} (\texttt{10v11}). 
In \texttt{3sv4z}, the allied team contains three Stalkers and the enemy team contains four Zealots. 
The remaining three scenarios use Marines, where the allied team contains five, eight, or ten Marines, respectively, and the enemy team contains one additional Marine. 
The action space is discrete and includes movement, attack actions targeting visible enemies, stop, and no operation, with invalid actions filtered by an action mask. 
Episodes terminate when all units of one side are eliminated or when a time limit is reached. 
Agents receive shaped rewards based on damage dealt, enemy eliminations, and a terminal reward for winning.
\begin{figure}[h]
	\centering
	\includegraphics[width=0.8\columnwidth]{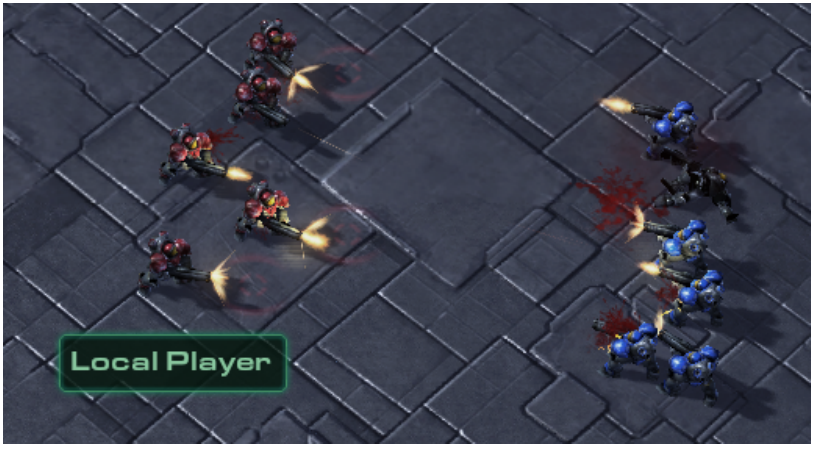}
	\caption{
		StarCraft Multi-Agent Challenge.}
	\label{fig:smac}
\end{figure}

\begin{table}[h]
	\centering
	\begin{tabular}{lccc}
		\toprule
		Map & Ally Units & Enemy Units & Steps \\
		\midrule
		3sv4z & 3 Stalkers & 4 Zealots & 20M \\
		5v6   & 5 Marines  & 6 Marines  & 20M \\
		8v9   & 8 Marines  & 9 Marines  & 20M \\
		10v11 & 10 Marines & 11 Marines & 20M \\
		\bottomrule
	\end{tabular}
	\caption{SMAC scenarios used in our experiments.}
\end{table}

\subsection{Hyperparameters}
\label{app:hparams}

Table~\ref{tab:hparams} summarizes the key hyperparameters used in our experiments. Unless otherwise stated, the same hyperparameters are used across all environments.

\begin{table}[t]
	\centering
	\begin{tabular}{p{0.58\columnwidth}p{0.22\columnwidth}}
		\toprule
		\textbf{Parameter} & \textbf{Value} \\
		\midrule
		Discount factor $\gamma$ & 0.99 \\
		GAE $\lambda$ & 0.95 \\
		Learning rate & $5\times10^{-4}$ \\
		Optimizer parameters & $\alpha=0.99$, $\epsilon=10^{-5}$ \\
		Entropy coefficient & 0.05 \\
		PPO clip $\epsilon$ & 0.1 \\
		Observation normalization & True \\
		Orthogonal initialization & True \\
		Advantage standardization & True \\
		Total environment steps & 20M \\
		\midrule
		Encoder embedding dim & 64 \\
		Encoder hidden dim & 64 \\
		Encoder hidden layers & 1 \\
		Contrastive projection dim & 32 \\
		Encoder--decoder epochs & 1 \\
		Encoder--decoder minibatches & 1 \\
		Encoder--decoder learning rate & $5\times10^{-4}$ \\
		\midrule
		Relational reasoning edge dim / global dim / hidden dim & 8 / 64 / 128 \\
		Relational reasoning message passing steps & 2 \\
		Relational reasoning aggregation operator & Sum \\
		\midrule
		Loss coefficients $\lambda_{\mathrm{AE}}$ / $\lambda_v$ & 1.0 / 0.5 \\
		Contrastive loss coefficient $\lambda_{\mathrm{con}}$ / temperature $\tau$ & 0.1 / 0.05 \\
		Contrastive sample cap & 512 \\
		\midrule
		Parallel environments & 8 \\
		Batch size / buffer size & 256 / 256 \\
		Evaluation episodes & 128 \\
		Random seeds & 5 \\
		\bottomrule
	\end{tabular}
	\caption{Key training hyperparameters used in our experiments.}
	\label{tab:hparams}
\end{table}

The PPO training settings differ slightly between POAM-MPG and IPPO-MPG: POAM-MPG uses $5$ optimization epochs per update with a single minibatch, whereas IPPO-MPG uses $4$ optimization epochs with $3$ minibatches per update and adopts the Huber loss for value learning. All other hyperparameters remain identical across algorithms. Unless otherwise noted, all experiments are implemented in EPyMARL using Sacred for configuration management, and reproducibility is ensured by setting the same random seed for NumPy, PyTorch, and the environment at the start of each run.\footnote{The codebase of EPyMARL is available at \url{https://github.com/uoe-agents/epymarl}.}

\subsection{PACT Implementation Details}
\label{app:pact_impl}

This subsection gives the implementation details used in our PACT runs. It is intended to make the configuration in Table~\ref{tab:hparams} unambiguous.

\paragraph{Agent modeling architecture.}The encoder is a GRUCell based recurrent encoder. At each timestep, its input is the concatenation of the local observation and the previous one hot action. The GRU hidden size is $64$. Its output is passed through one linear layer with ReLU activation and then through a linear embedding layer to produce a $64$ dimensional agent feature.

For each inferred target agent, the pair feature mapper takes the concatenation of the encoder feature and the one hot agent index. It is implemented as an MLP with one hidden layer, ReLU activation, hidden dimension $64$, and output dimension $64$. The decoder contains two parallel MLP heads with the same hidden dimension and activation. One head reconstructs the observation with mean squared error, and the other head predicts action logits optimized with cross entropy. The observation decoder uses a sigmoid output only when binary reconstruction is enabled, which is disabled in our reported experiments.

\paragraph{Contrastive projection and minibatch construction.}The contrastive projection head is a two layer MLP with ReLU activation, hidden dimension $64$, and output dimension $32$. The projected vectors are L2 normalized before cosine similarity is computed. In the released configuration, the supervised contrastive loss uses the phenotype label as its supervision signal. For uncontrolled teammates, this label is determined only by the reward shaping phenotype.

For each encoder--decoder update, let $\mathcal{J}_{\mathrm{unc}}$ denote the instantiated uncontrolled teammate slots in the sampled teams. The contrastive minibatch is constructed only from uncontrolled teammate slots with valid phenotype labels:
\[
\mathcal{Q}=\{(i,j,t): j \in \mathcal{J}_{\mathrm{unc}},\ y_{i,j}^t\ \mathrm{is\ valid},\ \mathrm{mask}_{i,j,t}=1\}.
\]
Controlled-agent slots are not treated as an additional phenotype class and are excluded from $L_{\mathrm{con}}$. This keeps the contrastive objective aligned with its role in Section~4.2: separating coordination phenotypes of uncontrolled teammate groups. The implementation keeps at most $512$ valid uncontrolled features and uses balanced sampling by phenotype label when this cap is reached. A valid feature must pass the episode mask. Therefore, padded timesteps and timesteps after episode termination are excluded. Agents that are not visible are not given a separate visibility mask, because visibility is already encoded in the SMAC observation and action availability tensors. With the default team mask, dead uncontrolled-agent slots are retained until the episode terminates. In settings where individual termination masks are enabled, dead agents are masked at the agent level through the available action based mask.

\paragraph{Relational reasoning module.}For each agent centered graph, nodes correspond to all agent slots in the team. The graph is a directed complete graph without self edges. Edge features have dimension $8$, node features have dimension $64$, and the global graph feature has dimension $64$. The relational reasoning module performs two message passing iterations, each consisting of an edge update, a node update, and a global update. All three update functions are two layer MLPs with ReLU activation and hidden dimension $128$. The aggregation operators from edges to nodes, edges to global features, and nodes to global features are sums.

\paragraph{Loss coefficients and optimization.}The agent modeling loss is the sum of the observation reconstruction loss and the action prediction loss, so $\lambda_{\mathrm{AE}}=1.0$. The contrastive term is added to the encoder--decoder loss with coefficient $\lambda_{\mathrm{con}}=0.1$. The value loss is optimized with coefficient $\lambda_v=0.5$, matching the critic update in the implementation. The relational reasoning parameters are shared by the actor and critic forward passes and are updated once after the actor and critic gradients have been accumulated.

\paragraph{Baseline adaptation.}IPPO-MPG, POAM-MPG, and PACT use the same controller for sampling teams, the same trainable agent mask for the actor, and the same evaluation protocol. IPPO-MPG removes the agent modeling, contrastive projection, and relational reasoning modules. POAM-MPG keeps the encoder--decoder agent modeling module, but removes the phenotype contrastive objective and the relational reasoning module. Thus, the comparison isolates the contribution of phenotype-aware contrastive learning and relational reasoning while keeping the MPG-AHT sampling process unchanged.

\subsection{Detailed Procedure for Sampling Teams}
\label{app:sampling}

Let $g(k,\mu)$ denote a sampling function that selects $k$ agents from a candidate set $\mu$. 
The function may sample with or without replacement depending on the environment configuration.

Let $\mathcal{U}=\{U_1,\dots,U_L\}$ denote the set of pre-trained uncontrolled teammate groups defined in Section~\ref{sec:problem_formulation}, and let $\mathcal{C}$ denote the set of controlled agents. 
Each group $U_\ell$ corresponds to one coordination phenotype induced by a fixed reward configuration. All agents within $U_\ell$ are trained jointly using the same MARL algorithm, so they have learned to coordinate with one another before being used in MPG-AHT. Different groups are trained independently, possibly using different MARL algorithms and reward configurations that induce different phenotypes.

For an environment containing $n$ agents, the stochastic \textit{procedure for sampling teams} $X$ follows the notation in Section~\ref{sec:problem_formulation}:
\[
\bm{\pi}^{(n)} \sim X(U_1,\dots,U_L, C(\theta)).
\]

The procedure is defined as follows:

\begin{enumerate}
	\item Sample the number of uncontrolled teammate groups $m$ and the group sizes
	\[
	m,(c_1,\dots,c_m,c_{\mathcal{C}}) \sim P_n,
	\]
	where $P_n$ is the distribution over valid group counts and sizes specified for each environment, $c_k$ is the number of agents in the $k$-th uncontrolled group, $c_{\mathcal{C}}$ is the number of controlled agents, and $\sum_{k=1}^{m} c_k + c_{\mathcal{C}} = n$.
	
	\item Sample $m$ distinct pre-trained groups
	\[
	U_{\ell_1},\dots,U_{\ell_m} \in \mathcal{U},
	\qquad
	\ell_k \neq \ell_{k'} \ \text{for}\ k\neq k'.
	\]
	In the multi-phenotype settings studied in this paper, the selected groups also have distinct phenotype labels. Each selected $U_{\ell_k}$ remains a jointly trained group rather than a collection of independently trained agents.
	
	\item Draw $c_k$ agents from each selected group $U_{\ell_k}$ using $g(\cdot,\cdot)$.
	
	\item Draw $c_{\mathcal{C}}$ controlled agents from $\mathcal{C}$ using $g(\cdot,\cdot)$.

	\item Combine the sampled uncontrolled agents and controlled agents into an $n$-agent team, and randomly assign the agents to the environment slots before deployment.
\end{enumerate}

This detailed procedure specifies how $X$ in Eq.~(\ref{eq:obj}) samples $\bm{\pi}^{(n)}$ for each episode, including the random assignment of the sampled agents to environment slots before deployment.

\subsection{Generation of Uncontrolled Teammate Groups with Different Phenotypes}
\label{app:phenotype_teammates}

To construct uncontrolled teammate groups with diverse phenotypes for MPG-AHT, we train groups using \emph{reward shaping conditioned on phenotype}. 
The key idea is to bias the reward structure during training so that agents learn distinct coordination tendencies. 
For each combination of a phenotype configuration and a MARL algorithm, we jointly train the agents of a separate uncontrolled teammate group. We use QMIX, IQL, VDN, IPPO, and MAPPO to construct such groups.
Thus, agents within a group are mutually familiar because they are trained together, whereas different groups are trained independently and have no prior experience collaborating with one another.
After group training, all agents in every uncontrolled teammate group are frozen. During MPG-AHT training and evaluation, agents are sampled from these intact pre-trained groups. Agents trained independently under different algorithms are never merged into a single group.

Figure~\ref{fig:two_stage_training} summarizes which policies are trainable or frozen in this procedure.
\begin{center}
	\centering
	\includegraphics[width=0.8\columnwidth]{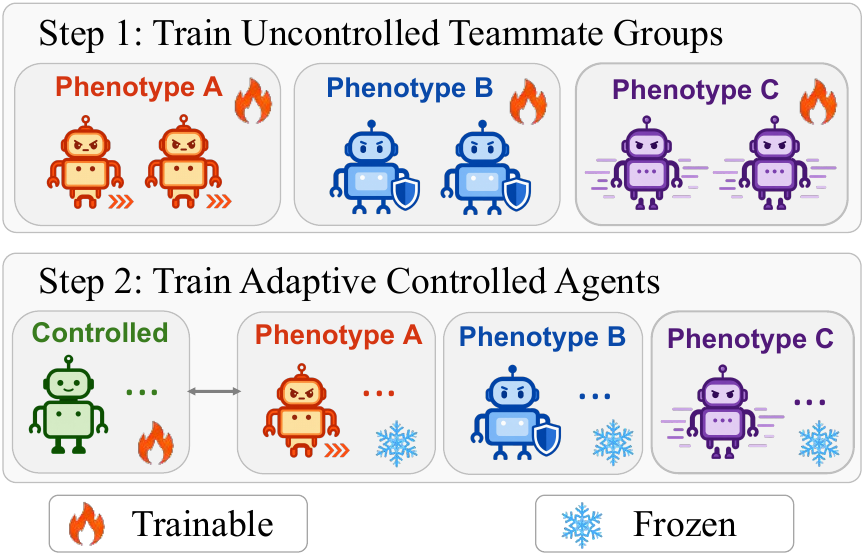}
	\captionof{figure}{
		Training procedure for MPG-AHT.
		In Step 1, the agents within each uncontrolled teammate group are trained jointly, while different groups are trained independently under their respective phenotype configurations.
		In Step 2, these pre-trained uncontrolled teammate groups are frozen, and only the controlled agents are optimized while interacting with teams composed of sampled groups.}
	\label{fig:two_stage_training}
\end{center}

Let $\mathcal{P}_{\mathrm{train}}$ and $\mathcal{P}_{\mathrm{test}}$ denote disjoint sets of phenotypes induced by reward shaping, and let $\mathcal{M}$ denote the set of MARL algorithms used to train uncontrolled teammate groups. For phenotype $p$ and algorithm $a$, $U_{p,a}$ denotes one jointly trained uncontrolled teammate group whose members share phenotype $p$. The training teammate pool and held-out OOD teammate pool are
\[
\begin{aligned}
U_{\mathrm{train}} &= \{U_{p,a} \mid p \in \mathcal{P}_{\mathrm{train}},\ a\in\mathcal{M}\}, \\
U_{\mathrm{test}} &= \{U_{p,a} \mid p \in \mathcal{P}_{\mathrm{test}},\ a\in\mathcal{M}\},\\
\mathcal{P}_{\mathrm{train}} &\cap \mathcal{P}_{\mathrm{test}} = \emptyset .
\end{aligned}
\]
Every $U_{p,a}$ is a separate frozen group: its agents were trained together using algorithm $a$, and agents from different values of $a$ are not pooled into the same group. During training, each episode samples teammate groups with distinct coordination phenotypes from $U_{\mathrm{train}}$, assigns the uncontrolled slots among the selected groups, and draws the required agents from each group. The remaining slots are filled by the controlled policy. The sampled groups together with the controlled agents form the team for that episode.

OOD evaluation uses the same team construction rule but draws uncontrolled teammates from $U_{\mathrm{test}}$. This keeps the number of uncontrolled agents, the team assembly process, and greedy evaluation unchanged while replacing the teammate phenotype pool with held-out phenotypes induced by reward shaping.

\paragraph{Behavioral Analysis of Phenotype Gaps.}The low, medium, and high phenotype-gap analysis in Figure~\ref{fig:phenotype_gap_robustness} uses behavioral distances between frozen uncontrolled teammate groups, not the reward shaping coefficients themselves.
For each uncontrolled teammate group $g$, we collect statistics from rollouts of its jointly trained, frozen members.
In SMAC, the raw behavior vector $\mathbf{x}^{\mathrm{SMAC}}_g$ comprises mean nearest enemy distance, first engagement fraction, attack rate, focus fire share, allied spatial dispersion, retreat rate at low health, move rate, stop rate, and the number of dead allied units.
In \texttt{mpe-pp}, the environment-specific raw behavior vector $\mathbf{x}^{\mathrm{MPE}}_g$ comprises, in order, contact rate, mean prey distance, mean team spread, mean predator speed, active rate, border pressure, and obstacle distance.
For each environment, every dimension is standardized over its corresponding candidate teammate group pool, yielding $\mathbf{b}^{e}_g$, where $e\in\{\mathrm{MPE},\mathrm{SMAC}\}$.
For two groups $g_i$ and $g_j$, we compute the phenotype gap as the root mean squared Euclidean distance between their standardized behavior vectors:
\[
d_{\mathrm{phen}}^{e}(g_i,g_j)
=
\sqrt{\frac{1}{K_e}
\sum_{k=1}^{K_e}
\left(b^{e}_{g_i,k}-b^{e}_{g_j,k}\right)^2},
\]
where $K_e$ denotes the number of behavior dimensions used for environment $e$.
The bucket boundaries are the 33rd and 67th percentiles of all pairwise distances within the corresponding candidate pool.
Table~\ref{tab:phenotype_gap_thresholds} reports the resulting thresholds for each environment.
Pairs with distances no larger than the low threshold are assigned to the low-gap bucket, pairs with distances larger than the high threshold are assigned to the high-gap bucket, and the remaining pairs are assigned to the medium-gap bucket.

\begin{table}[t]
	\centering
	\begin{tabular}{lcc}
		\toprule
		Environment & Low threshold & High threshold \\
		\midrule
		\texttt{mpe-pp} & 1.1629 & 1.6666 \\
		\texttt{3sv4z} & 1.1805 & 1.7118 \\
		\texttt{5v6} & 0.7141 & 1.0956 \\
		\texttt{8v9} & 0.6851 & 0.9879 \\
		\texttt{10v11} & 1.1721 & 1.5489 \\
		\bottomrule
	\end{tabular}
	\caption{Phenotype-gap bucket thresholds used in the robustness analysis.}
	\label{tab:phenotype_gap_thresholds}
\end{table}

\paragraph{Phenotype Generation in MPE Predator--Prey.}In the MPE Predator--Prey task, phenotype diversity is induced by modifying the aggregation of individual rewards returned by the environment. 
Let $r_{\text{pred}}^j$ denote the reward of predator $j$, and $r_{\text{prey}}$ the reward of the prey agent. 
We construct a team reward as

\begin{equation}
	r = \lambda_{\text{pred}} \sum_{j=1}^{3} r_{\text{pred}}^j + \lambda_{\text{prey}} r_{\text{prey}} .
\end{equation}

The coefficients $\lambda_{\text{pred}}$ and $\lambda_{\text{prey}}$ control the relative influence of predator capture rewards and feedback associated with the prey, thereby biasing the learned coordination strategies.

We define four phenotype variants:

\begin{itemize}
	\item \textbf{Phenotype A (strong pursuit phenotype):} $\lambda_{\text{pred}}=1.0$, $\lambda_{\text{prey}}=1.0$.
	\item \textbf{Phenotype B (standard pursuit phenotype):} $\lambda_{\text{pred}}=0.9$, $\lambda_{\text{prey}}=1.0$.
	\item \textbf{Phenotype C (mild risk-sensitive phenotype):} $\lambda_{\text{pred}}=0.7$, $\lambda_{\text{prey}}=1.0$.
	\item \textbf{Phenotype D (strong risk-sensitive phenotype):} $\lambda_{\text{pred}}=0.5$, $\lambda_{\text{prey}}=1.2$.
\end{itemize}

Intuitively, Phenotypes A--B encourage aggressive pursuit and faster capture, while Phenotypes C--D promote more cautious positioning and coordinated capture strategies.

\begin{figure*}[t]
	\centering
		\includegraphics[width=\textwidth]{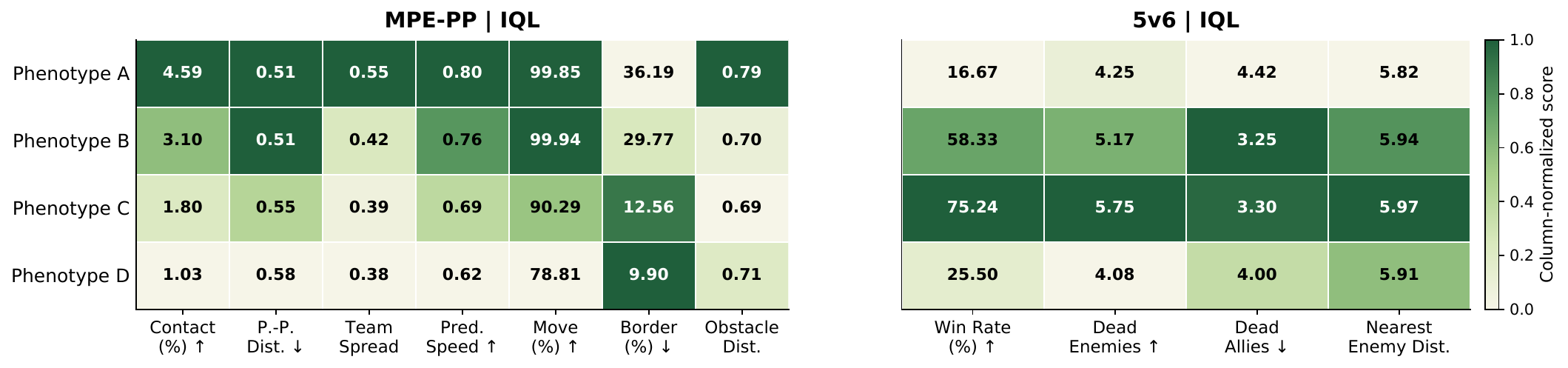}
	\caption{
		Left: behavior metrics of IQL teammates trained under four phenotypes on \texttt{mpe-pp}.
		Right: behavior metrics of IQL teammates trained under four phenotypes on SMAC \texttt{5v6}.
		Colors are normalized within each metric column, and numbers denote the raw values.}
	\label{fig:phenotype_behavior_heatmaps}
\end{figure*}

\begin{table*}[t]
	\centering
	\begin{tabular}{p{0.16\textwidth}p{0.21\textwidth}p{0.55\textwidth}}
		\toprule
		Environment & Metric & Description \\
		\midrule
		\texttt{mpe-pp}
		& Contact rate
		& Percentage of evaluated timesteps in which at least one predator is in contact with the prey. \\
		& Predator--prey distance
		& Average distance from predators to the prey. Lower values indicate closer pursuit. \\
		& Team spread
		& Average pairwise distance among predators, reflecting how dispersed the pursuing team is. \\
		& Predator speed
		& Average movement speed of predators during pursuit. \\
			& Active rate
		& Fraction of agent decisions assigned to movement actions rather than staying still. \\
		& Border pressure
		& Frequency with which predators move close to the environment boundary. \\
		& Obstacle distance
		& Average distance from each predator to its nearest obstacle. \\
		\midrule
		SMAC \texttt{5v6}
		& Win rate
		& Fraction of evaluation episodes won by allied units. \\
		& Dead enemies
		& Average number of enemy units eliminated. \\
		& Dead allies
		& Average number of allied units lost. \\
		& Nearest enemy distance
		& Average distance from each allied unit to its closest visible enemy. \\
		\bottomrule
	\end{tabular}
		\caption{Behavior metrics used in Figure~\ref{fig:phenotype_behavior_heatmaps}.}
		\label{tab:phenotype_behavior_metrics}
\end{table*}

Since our experiments cover both MPE and SMAC, Figure~\ref{fig:phenotype_behavior_heatmaps} uses one representative task from each environment to visualize how phenotypes induced by reward shaping translate into different teammate behaviors.
The left panel of Figure~\ref{fig:phenotype_behavior_heatmaps} visualizes how these phenotype choices are reflected in IQL teammates on \texttt{mpe-pp}, using the seven MPE metrics summarized in Table~\ref{tab:phenotype_behavior_metrics}.
Together, these metrics provide a compact behavioral view of pursuit strength, caution, movement intensity, and spatial coordination, which helps verify that the reward configurations induce visibly different coordination phenotypes among teammates in the MPE environment.

\paragraph{Phenotype Generation in SMAC.}In SMAC, we introduce phenotype variations by modifying the dense reward with an additional distance shaping term. 
Let $r_{\text{SMAC}}$ denote the original dense reward composed of damage, kill bonus, and win bonus. 
The final reward is defined as

\begin{equation}
	r = r_{\text{base}} + r_{\text{dist}}, 
	\qquad 
	r_{\text{dist}} = \beta \sum_{i=1}^{M} d_i ,
\end{equation}

where $M$ is the number of allied agents and $d_i$ is the distance between agent $i$ and its closest visible enemy.

For each agent $i$, the closest enemy distance is computed as

\[
d_i = \min_{e \in \mathcal{E}_i} \| \Delta \mathbf{p}_{i,e} \|_2 ,
\]

where $\mathcal{E}_i$ is the set of visible enemies and $\Delta \mathbf{p}_{i,e}$ denotes the relative position of enemy $e$ in the observation.

The base reward $r_{\text{base}}$ follows the SMAC reward structure. 
Let $r_{\text{damage}}$ and $r_{\text{kill}}$ denote rewards for damaging and eliminating enemies, and $r_{\text{win}}$ the episode win bonus. 
When negative rewards are enabled, a penalty term $r_{\text{ally}}$ captures allied damage or deaths:

\begin{equation}
	r_{\text{base}} =
	\begin{cases}
		\left|\alpha (r_{\text{damage}} + r_{\text{kill}})\right| + r_{\text{win}},
		& \text{positive reward}, \\
		\begin{aligned}
			&\alpha (r_{\text{damage}} + r_{\text{kill}})\\
			&\quad - \eta r_{\text{ally}} + r_{\text{win}},
		\end{aligned}
		& \text{penalized reward}.
	\end{cases}
\end{equation}

Using this formulation, we instantiate four teammate phenotypes:

\begin{itemize}
	\item \textbf{Phenotype A (strong attack phenotype):} positive reward setting, $\alpha=1.2$, $\beta=-0.5$.
	\item \textbf{Phenotype B (standard attack phenotype):} positive reward setting, $\alpha=1.0$, $\beta=0$.
	\item \textbf{Phenotype C (mild risk-sensitive phenotype):} penalized setting, $\alpha=1.0$, $\eta=0.2$, $\beta=+0.5$.
	\item \textbf{Phenotype D (strong risk-sensitive phenotype):} penalized setting, $\alpha=1.0$, $\eta=0.5$, $\beta=+1.0$.
\end{itemize}

Phenotypes A--B emphasize aggressive enemy elimination, while Phenotypes C--D introduce penalties and distance incentives that encourage more cautious behavior.

The right panel of Figure~\ref{fig:phenotype_behavior_heatmaps} reports the four SMAC metrics summarized in Table~\ref{tab:phenotype_behavior_metrics}.
Several qualitative differences can be observed.
Phenotype C achieves the highest win rate and the largest number of dead enemies, while also producing the fewest allied deaths.
It also has the largest mean nearest enemy distance, which is consistent with a more cautious phenotype that still maintains effective combat performance.
By contrast, Phenotype A yields more allied deaths and a lower win rate, indicating a more aggressive but less stable behavior pattern.
Phenotypes B and D fall between these extremes, further showing that the proposed reward shaping produces meaningfully different coordination tendencies even within the same learning algorithm.

\subsection{Evaluation Procedure}
\label{app:eval_procedure}

\begin{figure*}[t]
	\centering
	\includegraphics[width=\textwidth]{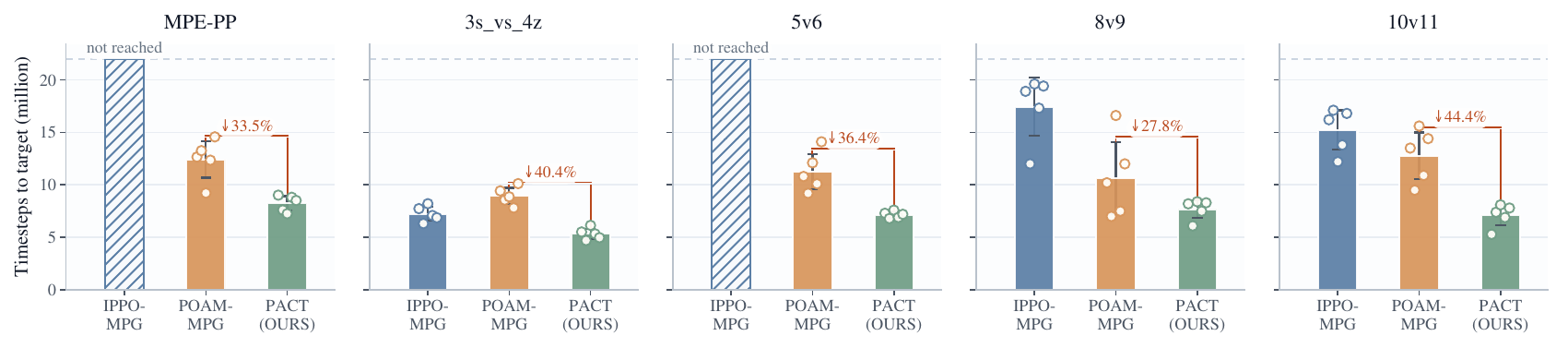}
	\caption{
		Sample efficiency measured by the number of timesteps required to reach the target score on each task. The target score is the strongest naive MARL baseline in the corresponding environment, and hatched bars indicate that the target is not reached within the training budget of $20$M environment steps.}
	\label{fig:sample_efficiency_thresholds}
\end{figure*}
\begin{figure*}[h]
	\centering
	\includegraphics[width=\textwidth]{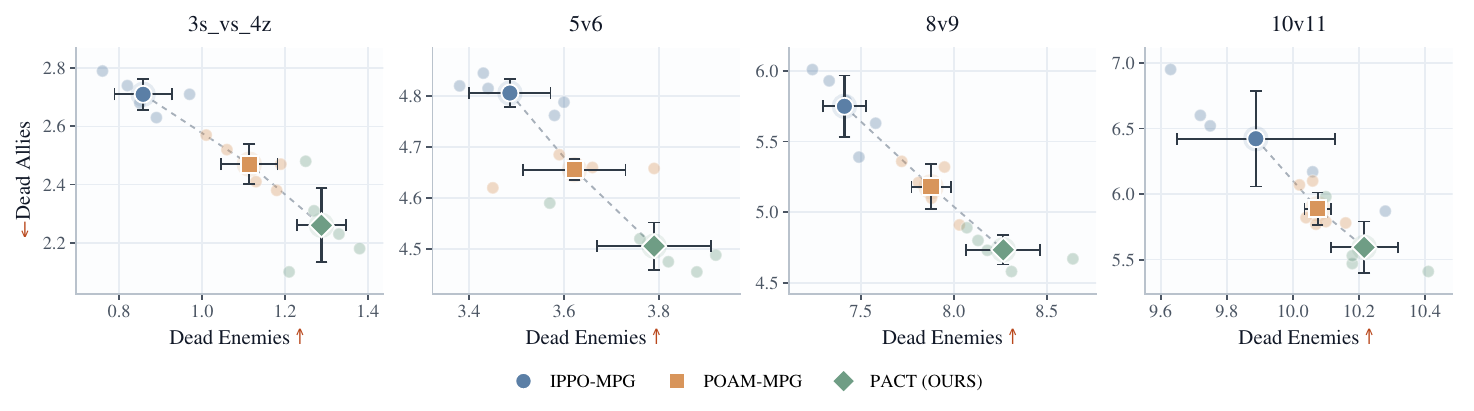}
	\caption{
		Battle outcome tradeoff on SMAC under OOD evaluation. Each point denotes one seed, and the marker with error bars denotes the mean and $95\%$ confidence interval. Better performance lies toward the lower right, corresponding to more dead enemies and fewer dead allies.}
	\label{fig:battle_tradeoff}
\end{figure*}

We evaluate the controlled agents by constructing mixed groups consisting of the trained policy and a set of uncontrolled teammates sampled from pre-trained groups associated with specific phenotypes.
For each evaluation configuration, the number of uncontrolled agents is fixed to $n_{\mathrm{unc}}$, and we sweep $n_{\mathrm{unc}}$ from $2$ to $M-1$ to evaluate performance under different team compositions, where $M$ denotes the total number of agents in the environment.
For each value of $n_{\mathrm{unc}}$, evaluation is performed over $128$ episodes using greedy action selection.
Unless otherwise specified, final results first average the logged statistics within each seed over evaluation episodes, swept values of $n_{\mathrm{unc}}$, and sampled combinations of held-out teammate groups. The reported mean and $95\%$ confidence interval are then computed across the estimates for the five seeds.

At the beginning of each evaluation episode, the number of uncontrolled teammate groups follows the evaluation setting: the main experiments use two uncontrolled groups, while the Random-N-Phenotype experiment samples this number from $\{2,3,4\}$.
The $n_{\mathrm{unc}}$ uncontrolled agents are then partitioned among the selected groups according to the team-sampling procedure $X$.
The specific uncontrolled agents are drawn without replacement from the corresponding pre-trained groups, while the remaining slots are filled by the controlled policy.

\paragraph{OOD Evaluation.}For out-of-distribution evaluation, the sampling procedure remains identical, but the selected uncontrolled teammate group or groups are drawn from $U_{\mathrm{test}}$ rather than $U_{\mathrm{train}}$.
The set $U_{\mathrm{test}}$ contains groups trained with reward shaping phenotypes that are unseen during training.
By keeping the team construction process unchanged and altering only the source of teammates, this protocol isolates the effect of phenotype shift among teammates and directly measures generalization to unseen coordination behaviors.
The mean OOD performance gain of $21.0\%$ reported in the main text is computed from Table~\ref{tab:maht_win} by comparing PACT with the strongest baseline on each task and then averaging the relative gains across the five tasks.

\subsection{Computational Infrastructure}
\label{app:computing_infra}

Value-based methods (e.g., QMIX~\cite{qmix}, VDN~\cite{vdn}, and IQL~\cite{iql}) were trained using rollouts from one environment instance, while policy gradient methods (e.g., IPPO-MPG, POAM-MPG, and PACT) used parallel environment rollouts.

All experiments were conducted on a workstation running Ubuntu 20.04 with an AMD Ryzen CPU and an NVIDIA RTX 4090 GPU.

\section{Supplementary Experimental Results}
\label{app:supplementary_experimental_results}

\subsection{Hyperparameter Sensitivity}
\label{app:hyperparameter}

We analyze sensitivity to the contrastive loss coefficient $\lambda_{\mathrm{con}}$ and temperature $\tau$ on \texttt{5v6}.
For the coefficient study, we fix $\tau=0.05$ and vary $\lambda_{\mathrm{con}} \in \{0.05, 0.1, 0.2, 0.4, 0.8\}$.
For the temperature study, we fix $\lambda_{\mathrm{con}}=0.1$ and vary $\tau \in \{0.03, 0.05, 0.1\}$.

\begin{center}
	\centering
	\includegraphics[width=\columnwidth]{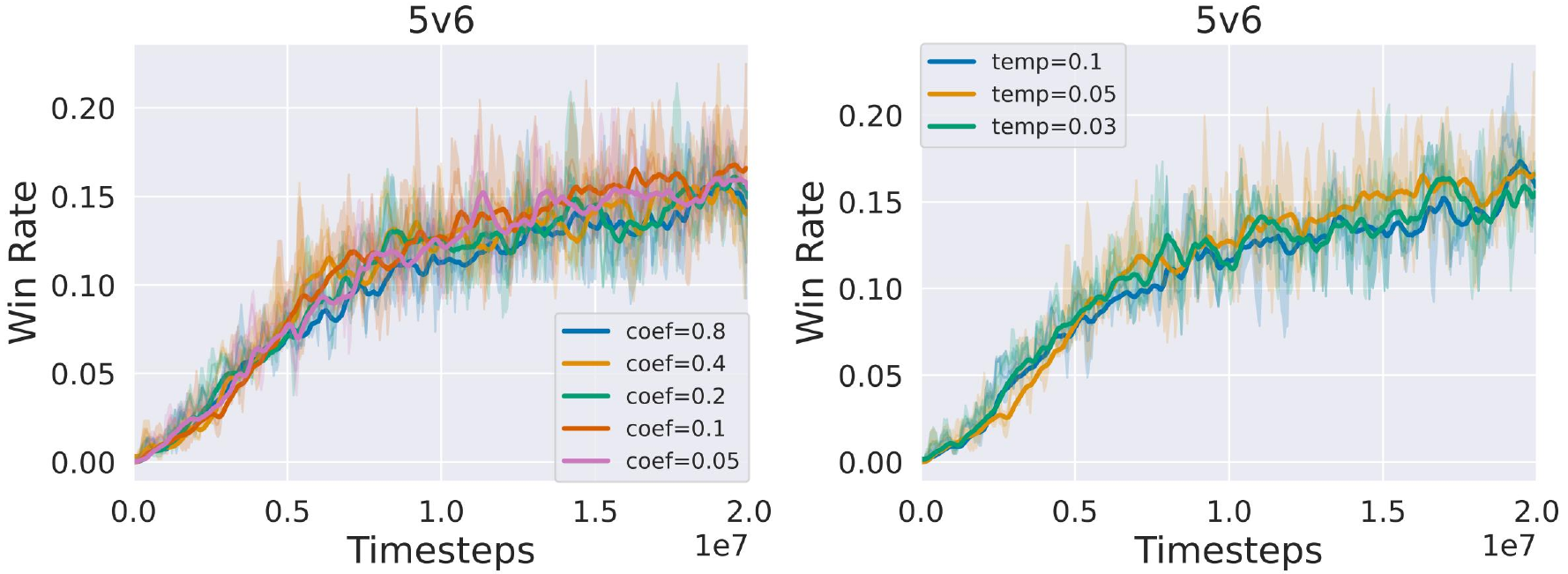}
	\captionof{figure}{
		Hyperparameter sensitivity of the contrastive objective on \texttt{5v6}.
		The two panels vary the contrastive loss coefficient $\lambda_{\mathrm{con}}$ and the temperature $\tau$, respectively.}
	\label{fig:hyperparameter_sensitivity}
\end{center}

As shown in Figure~\ref{fig:hyperparameter_sensitivity}, PACT remains stable across a broad range of hyperparameter values, with $\lambda_{\mathrm{con}}=0.1$ and $\tau=0.05$ achieving the best performance.
Therefore, these values are used as the default hyperparameters in all experiments.

\begin{figure*}[h]
	\centering
	\includegraphics[width=1.0\textwidth]{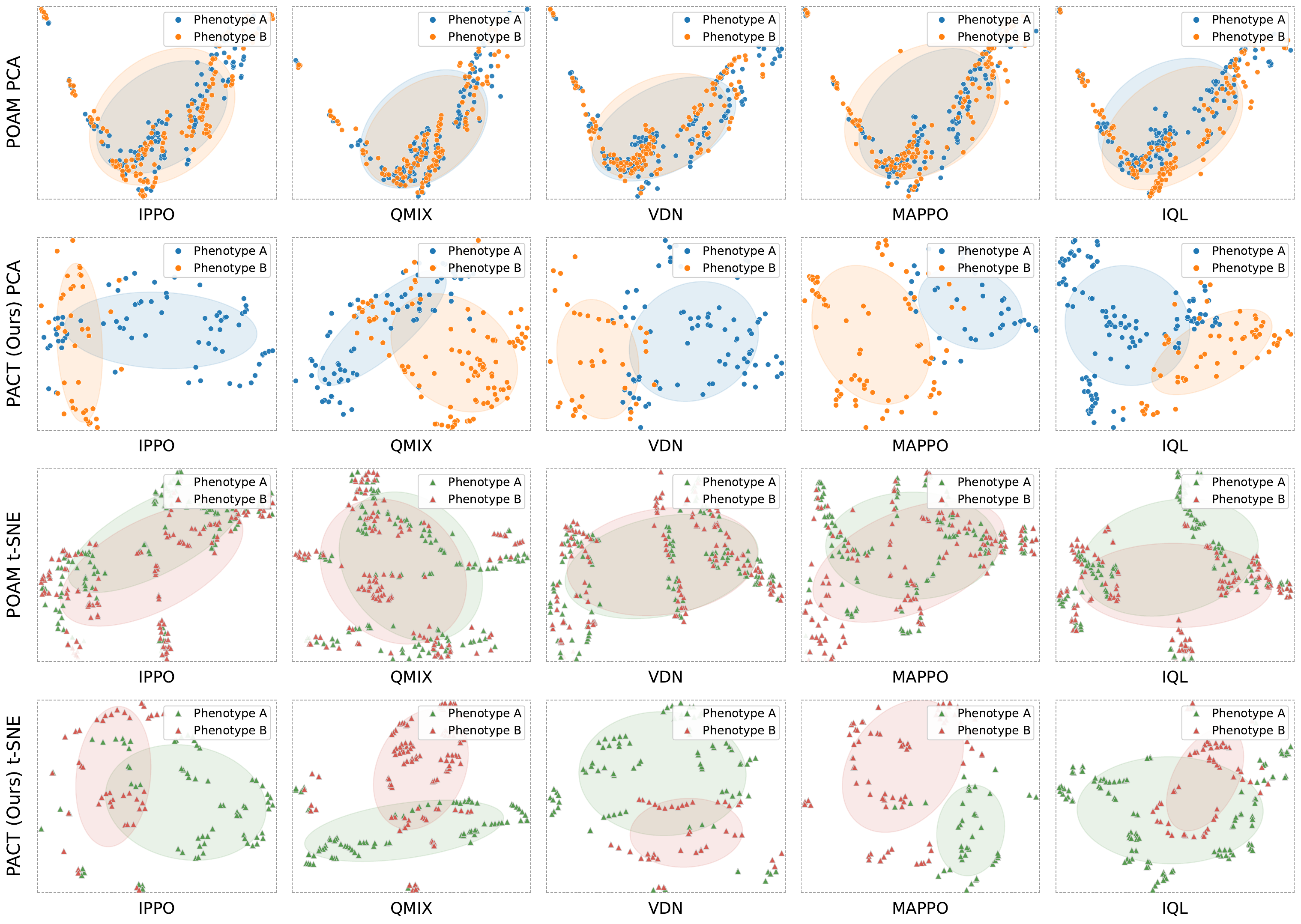}
	\caption{
		Complete PCA and t-SNE visualization of POAM encoder features and PACT projection features under phenotypes A and B.
		Compared with POAM encoder features, PACT projection features show consistently clearer phenotype separation across algorithms.}
	\label{fig:contrastive_remaining_vis}
\end{figure*}

\subsection{Additional Analysis of Training Efficiency and Battle Outcomes}
\label{app:training_battle_analysis}

To complement the learning curves in Figure~\ref{fig:main_result1}, we summarize sample efficiency by measuring, for each seed, the first timestep at which the evaluation performance reaches a target selected for each task.
For each environment, the target score is defined as the test return or win rate achieved by the strongest naive MARL baseline, corresponding to the dashed reference line in the main paper.
Figure~\ref{fig:sample_efficiency_thresholds} reports the mean and $95\%$ confidence interval across five seeds.

We quantify the sample-efficiency gain relative to POAM-MPG, which reaches the target on all five tasks, by the relative reduction in the number of environment steps: $100\%\times(1-T_{\mathrm{PACT}}/T_{\mathrm{POAM}})$ for each task. The reductions are $33.5\%$, $40.4\%$, $36.4\%$, $27.8\%$, and $44.4\%$ on \texttt{mpe-pp}, \texttt{3sv4z}, \texttt{5v6}, \texttt{8v9}, and \texttt{10v11}, respectively, yielding a mean reduction of $36.5\%$ over the five tasks. This is a relative reduction in required environment steps, rather than an absolute performance improvement.

PACT reaches the target score faster than POAM-MPG on all five tasks and faster than IPPO-MPG wherever both methods reach the target within the training horizon.
On \texttt{mpe-pp} and \texttt{5v6}, IPPO-MPG never reaches the target score within the budget of $20$M environment steps, whereas PACT does so consistently.
The advantage is also clear on the larger SMAC maps \texttt{8v9} and \texttt{10v11}, where PACT requires substantially fewer timesteps than both baselines.
These results provide a more direct quantitative view of the training efficiency gains of PACT under multi-phenotype collaboration.

We further analyze the final battle outcomes on SMAC by jointly plotting the number of dead enemies and dead allies in Figure~\ref{fig:battle_tradeoff}.
This view complements win rate by revealing whether performance gains arise from more effective combat or merely from riskier exchanges.

Across all four SMAC scenarios, PACT is consistently shifted toward the lower right relative to IPPO-MPG and POAM-MPG, indicating that it defeats more enemies while losing fewer allied units.
The separation is especially pronounced on \texttt{5v6}, \texttt{8v9}, and \texttt{10v11}, where PACT improves both axes simultaneously.
This suggests that the stronger performance of PACT reflects better coordination quality with unfamiliar teammates rather than a simple increase in aggressiveness.

\subsection{Additional Visualization of Phenotype Separation}
\label{app:phenotype_visualization}

Figure~\ref{fig:contrastive_remaining_vis} complements the visualization in the main paper by showing both PCA results on the remaining algorithms and t-SNE results across all five teammate algorithms.
Across these algorithms, POAM encoder features remain largely entangled across phenotypes, whereas PACT projection features show more compact groups for the same phenotype and clearer separation across phenotypes.
The comparison is intentionally made between POAM encoder features and PACT projection features under the same underlying teammate algorithms, so the observed separation is not explained by changing the teammate pool or the visualization protocol. Instead, it reflects how the phenotype-aware contrastive objective reshapes the representation after agent modeling.

In the POAM panels, samples from phenotypes A and B often occupy overlapping regions, especially when the underlying MARL algorithm already induces broad behavioral variation within a phenotype. This indicates that reconstructing observations and predicting actions can encode useful teammate behavior while still mixing coordination tendencies specific to each phenotype in the latent space. By contrast, the PACT panels show more separated clusters under both PCA and t-SNE, suggesting that the projection head organizes teammates according to the phenotype labels used by the contrastive objective.

The t-SNE rows provide a nonlinear view that complements the PCA rows. Although t-SNE should not be interpreted as preserving exact global distances, the repeated separation pattern across IPPO, QMIX, VDN, MAPPO, and IQL indicates that the effect is not tied to one particular teammate training algorithm. The consistency across both visualization methods strengthens the claim that PACT learns phenotype sensitive features rather than merely producing an artifact of a single visualization method.

These visualization results also help explain the empirical trends in the OOD and phenotype gap experiments. When teammate phenotypes are held out or paired with unfamiliar groups, the controlled agents must condition their behavior on coordination tendencies that are not captured by a single teammate identity or algorithm label. A representation that keeps phenotype information more explicit can therefore provide a more stable input to relational reasoning and policy learning, which is consistent with PACT's stronger generalization and robustness in the main results.

Taken together, Figure~\ref{fig:contrastive_remaining_vis} provides qualitative evidence for the role of the contrastive projection rather than serving as an additional performance metric. The goal is not to claim that the two phenotypes become perfectly linearly separable in every algorithm panel. Instead, the important observation is that PACT repeatedly reduces the overlap between phenotype groups compared with the corresponding POAM encoder features. This is the behavior expected from Eq.~(\ref{eq:contrastive}), which uses phenotype labels to pull samples from the same phenotype together while reducing the relative similarity of samples from different phenotypes.

The remaining overlap in several PACT panels is also informative. It suggests that the learned representation does not discard all variation caused by the algorithm or by individual trajectories, which would be undesirable because teammates with the same phenotype can still differ in execution details. PACT therefore appears to preserve behavioral information needed for action prediction while making phenotype structure more accessible to the downstream relational module. This interpretation is consistent with the ablation results in the main paper: contrastive learning improves representation structure, while relational reasoning converts that structured teammate information into better coordination decisions.

\end{document}